\RequirePackage{etex}
\documentclass[11pt,a4paper]{article}
\usepackage[T1]{fontenc}
\usepackage[left=1in, right=1in, vmargin=1in]{geometry}
\usepackage[english]{babel}
\usepackage[nospace,english]{varioref}
\labelformat{section}{section~#1}
\labelformat{figure}{figure~#1}
\usepackage{microtype}
\usepackage{amssymb}
\usepackage{mleftright}
\mleftright % \ remplace les \left et \right du code par \mleft et \mright
\usepackage{mismath}
\pinumber % pi droit
\usepackage[capsit]{frenchmath} % capsit : on conserve les majuscules en italique
\usepackage{pm-isomath} % fournit les majuscules grecques en italique
\switchvarlowercasegreekletters

\usepackage{csquotes}

\usepackage{yhmath} % pour l'arc \wideparen
\usepackage{mathrsfs}
\usepackage{siunitx}
\DeclareSIUnit\parsec{\text{pc}}
\usepackage{enumitem}
\usepackage{revsymb} % fourni \lambdabar
\makeatletter
\newcommand{\lambdabarEgreg}{{\mathchoice
  {\smash@bar\textfont\displaystyle{0.25}{1.2}\lambda}
  {\smash@bar\textfont\textstyle{0.25}{1.2}\lambda}
  {\smash@bar\scriptfont\scriptstyle{0.25}{1.2}\lambda}
  {\smash@bar\scriptscriptfont\scriptscriptstyle{0.25}{1.2}\lambda}
}}
\newcommand{\smash@bar}[4]{%
  \smash{\rlap{\raisebox{-#3\fontdimen5#10}{$\m@th#2\mkern#4mu\mathchar'26$}}}%
}
\makeatother

\usepackage{graphicx}
\usepackage{soul}
\usepackage{marginnote}

\usepackage{alphalph}
\newcounter{notemarginale}

\usepackage{pgf-pie}

\makeatletter
\def\pgfpie@numbertext#1{% don't print percentage in slice for small values
  \pgfpie@ifhidenumber{}{%
    \pgfmathparse{#1 > 5}%
    \ifnum\pgfmathresult=1 %
    \pgfpie@beforenumber\color{white}#1\pgfpie@afternumber%
    \else%
    \pgfpie@beforenumber#1\pgfpie@afternumber%
    \fi
  }%
}
\makeatother

\usepackage{worldflags}
\flagsdefault[framewidth=0.1pt,framecolor=black]
\usepackage[autolang=other*, maxnames=4,sorting=nyt]{biblatex} % prend en compte la langue de l'entrée lorsqu'un champ "hyphenation = {american}" est présent (apparement la langue "english" n'est pas reconnue) ; cette langue doit être présente dans le chargement de babel ; définit le nombre maxi de noms du champ auteur affichés avant d'utiliser et al 
\bibliography{LegacySakharov}

\usepackage[colorlinks,allcolors=blue]{hyperref}
\usepackage{orcidlink} % après hyperref, sinon ça produit un bug

\newcommand{\transp}[1]{#1^{\mathsf{T}}} % transposée

\newcommand{\Loo}{L_\text{o}}
\newcommand{\Rmunu}{R_{\mu\nu}}
\newcommand{\Rbarmunu}{\overline{R}_{\mu\nu}}
\newcommand{\gmunu}{g_{\mu\nu}}
\newcommand{\gbarmunu}{\overline{g}_{\mu\nu}}
\newcommand{\Tmunu}{T_{\mu\nu}}
\newcommand{\Tbarmunu}{\overline{T}_{\mu\nu}}
\newcommand{\Rbar}{\overline{R}}
\newcommand{\Sbar}{\overline{S}}

\newcommand{\gbar}{\lbar{g}}
\newcommand{\moins}{{(-)}}
\newcommand{\plus}{{(+)}}

\newcommand{\Rmixmunu}{R_{\mu}^{\nu}}
\newcommand{\gmixmunu}{g_{\mu}^{\nu}}
\newcommand{\Tmixmunu}{T_{\mu}^{\nu}}
\newcommand{\tmixmunu}{\mathcal{T}_{\mu}^{\nu}}
\newcommand{\Rbarmixmunu}{\overline{R}_{\mu}^{\nu}}
\newcommand{\gbarmixmunu}{\lbar{g}_{\mu}^{\nu}}
\newcommand{\Tbarmixmunu}{\overline{T}_{\mu}^{\nu}}
\newcommand{\tbarmixmunu}{\overline{\mathcal{T}}_{\mu}^{\nu}}
\newcommand{\abar}{\lbar{a}}
\newcommand{\kbar}{\lbar{k}}
\newcommand{\cbar}{\lbar{c}}

\newcommand{\rhobar}{\lbar{\rho}}

\newcommand{\upnablabar}{\overline{\nabla}}
\newcommand{\bbar}{\hlbar{b}}
\newcommand{\Tmixzz}{T_0^0}
\newcommand{\tmixzz}{\mathcal{T}_0^0}
\newcommand{\Tbarmixzz}{\overline{T}_0^0}
\newcommand{\tbarmixzz}{\overline{\mathcal{T}}_0^0}
\newcommand{\tJbar}{\lbar{t}_\text{J}}
\newcommand{\vbar}{\lbar{v}}
\newcommand{\pbar}{\lbar{p}}
\newcommand{\Rs}{R_\text{S}}
\newcommand{\nubar}{\lbar{\nu}}
\renewcommand{\lambdabar}{\lbar{\lambda}}% renewcommand, mais à voir si on vire la définition ou la redéfinition.
\newcommand{\sbar}{\lbar{s}}

\newcommand{\Mbar}{\lbar{M}}

\newcommand{\rzbar}{\lbar{r}_0}
\usepackage{orcidlink}
\usepackage{authblk}
\makeatletter
\newcommand{\liste}[1]{%
	\@tempcnta=0 % Compteur temporaire
	\def\@sep{}% Définir le séparateur initial (vide)
	\@for\@tempa:=#1\do{%
		\advance\@tempcnta by 1 % Incrément du compteur
		\@sep \@tempa\def\@sep{,\,}% Ajouter l'élément avec le séparateur
	}%
}
\makeatother

\begin{document}

\title{A bimetric cosmological model based on\\ Andreï Sakharov's twin universe approach}
\author{\small Jean-Pierre Petit\,\orcidlink{0000-0003-3141-8584}\thanks{\href{mailto:jean-pierre.petit@manaty.net}{jean-pierre.petit@manaty.net}}}
\affil{\itshape Manaty Research Group}

\author{\small Florent Margnat\thanks{\href{mailto:florent.margnat@univ-poitiers.fr}{florent.margnat@univ-poitiers.fr}}}
\affil{\itshape University of Poitiers}

\author{\small Hicham Zejli\,\orcidlink{0009-0006-8886-7101}\thanks{\href{mailto:hicham.zejli@manaty.net}{hicham.zejli@manaty.net}}}
\affil{\itshape Manaty Research Group}

\date{}

\maketitle

\tableofcontents

\vfill

%\textbf{Keywords :} Sakharov, twin universe, antimatter, dynamic groups theory, momentum, fifth dimension, topology, biometric model, Janus, T-symmetry, PT-symmetry, CPT-symmetry, Poincaré group, orthochron subgroup, antichron subset, \textit{Janus group}, Kaluza space, two-folds cover of a manifold, negative mass, coupled field equations system, runaway phenomenon, action, enantiomorphy, dipole repeller, dark matter, dark energy, very large structure, Newtonian approximation.

\newpage 

\begin{abstract}
The standard cosmological model, based on Cold Dark Matter and Dark Energy ($\Lambda$CDM), faces several challenges. Among these is the need to adjust the scenario to account for the presence of vast voids in the large-scale structure of the universe, as well as the early formation of the first stars and galaxies. Additionally, the observed matter-antimatter asymmetry in the universe remains an unresolved issue. To address this latter question, Andrei Sakharov proposed a twin universe model in 1967. Building upon this idea and introducing interactions between these two universe sheets through a bimetric model, we propose an alternative interpretation of the large-scale structure of the universe, including its voids and the acceleration of cosmic expansion.
\end{abstract}

\section{Introduction}

Between 1967 and 1980, the physicist Andreï Sakharov published several papers (\cite{Sakharov1967, Sakharov1979, Sakharov1980}) in which he presented a cosmological model with two universes, connected by an initial singularity: the Big Bang. The first universe corresponds to ours, while the second is described by Sakharov as a twin universe. The \textit{"arrows of time"} of these two universes are antiparallel, and they are \textit{"enantiomorphic"}, that is, mirrored. Through this model, Sakharov proposed a possible explanation for the apparent absence of primordial antimatter in our universe.\\

For more than half a century, cosmology has been unable to solve one of its greatest enigmas: not only has no convincing explanation been found as to why one particle of matter in a million escaped total annihilation with antimatter, but no significant observation of a corresponding amount of primordial antimatter has been made.\\

Sakharov was interested in the violation of CP-symmetry, a fundamental property of the laws of physics, and hypothesized that a twin universe, where these violations would be reversed, could exist. This model would thus restore a generalized symmetry on a large scale. Based on the fact that matter is formed from the assembly of quarks and antimatter from antiquarks, he supposed that in our universe, the reaction leading to the formation of matter would have been slightly faster than the one leading to the formation of antimatter, while the inverse situation would occur in the twin universe.\\

Thus, in our matter-dominated universe, there would remain a small surplus of matter, accompanied by an equivalent amount of free antiquarks. Symmetrically, in the twin universe, one would find antimatter with a corresponding surplus of free quarks. Although this model may seem exotic, it nevertheless offers the only theoretical explanation proposed so far to account for the disappearance of half of the predicted cosmic content. Consequently, it seems legitimate to examine in detail the aspects and implications of such a model.\\

This article revisits the pioneering work of Andreï Sakharov and proposes a new cosmological model, inspired by his approach, in which two folds of the universe are connected by the same initial singularity, folded over one another and interacting through gravitational effects. It puts this work in perspective with modern concepts to address some of the challenges posed by the standard cosmological model, particularly those of the $\Lambda$CDM model. This model offers potential explanations for phenomena such as the acceleration of cosmic expansion or the existence of large-scale structures like cosmic voids.\\

Our paper is structured around several key sections. The first explores T-symmetry, which corresponds to time reversal, based on the mathematical framework of the Poincaré group. This symmetry is related to the existence of particles with negative mass and energy, at the core of the bimetrical Janus model, inspired by the work of J.-M. Souriau. It plays a central role in the dynamics of this double spacetime, where time reversal opens the door to a new interpretation of physical phenomena (\cite{desaxce2023presentation, Souriau1970structure}).\\

Next, C-symmetry, associated with charge conjugation, is extended within the framework of an additional dimension through the Kaluza-Klein model. This extension allows the interpretation of electric charge as a geometric component, in accordance with Noether's theorem. This connection between the extra dimension and charge conservation offers a new perspective on charged particles in a five-dimensional spacetime, where charge naturally emerges from geometry (\cite{Souriau1970structure}).\\

The model is enriched by the introduction of the \textit{Janus restricted group}, which extends the Kaluza space to several compactified dimensions. This dynamic group links the matter-antimatter symmetry (C-symmetry) to the inversion of quantum charges in a multidimensional framework. Through this extension, the group's geometry allows for the understanding of the quantization of several charges, including electric charge, and opens the way to the emergence of new quantum charges. This section establishes a connection between Souriau's work and the Kaluza-Klein formalism to explain complex physical phenomena in a higher-dimensional spacetime (\cite{Kerner, desaxce2024symmetry, SouriauGeomRelat, Souriau1970structure}).\\

The \textit{Janus dynamic group}, which combines PT-symmetry (simultaneous inversion of energy, time, and spatial coordinates) and C-symmetry (charge conjugation), allows the modeling of interactions between matter, antimatter, and negative mass particles. Thanks to Noether's theorem, this group associates scalar invariants with the observed symmetries, thus clarifying the interactions between these different entities within a bimetrical framework, and allowing the extension of Sakharov's model by adding compactified dimensions for each quantum charge (\cite{Petit2014, Sakharov1967, Souriau1970structure}).\\

To illustrate this concept, in the context of our study on bimetric models, we proposed a model of wormhole linking two {PT-symmetric} folds of the universe via a modified Einstein-Rosen bridge~\cite{aop2024}. This model includes a cross term \(dr \, dt\) in the corresponding metric, leading to a finite free-fall time to the wormhole’s throat for an external observer. The two folds are {CPT-symmetric} for photons, which are neutral particles. This wormhole model allows for unidirectional traversal through its throat, inducing a space-time inversion. This opens the possibility of interactions between matter and antimatter, arising from the {PT} symmetry observed during the transition between the two universe folds. Thus, the congruent identification of points on the two universe folds and the reversal of the arrow of time induce an inversion of energy, offering new insights into the structure of space-time and the potential inversion of particle mass while crossing this bridge.\\

The Janus model will also be studied from a topological perspective, with a closed universe geometry where P and T symmetries naturally emerge. Spacetime is modeled by a compact universe with the topology of a 4-dimensional sphere \(\mathbb{S}^4\), which forms a two-fold cover of the projective space \(\mathbb{P}^4\). In this structure, the antipodal points, representing the Big Bang and the Big Crunch, coincide. By replacing these singularities with a tubular structure, they disappear, allowing P and T symmetries to emerge as natural consequences of this closed projective geometry (\(\mathbb{P}^4\)) and be interpreted in a purely topological framework (\cite{Boy1903, Petit1994, Petit-Topologicon}).\\

One of the earliest attempts to introduce negative masses into a cosmological model, explored by H. Bondi in 1957, showed that the coexistence of positive and negative masses, which respectively induce attraction and repulsion, leads to the \textit{"runaway effect"} (\cite{Bondi}). In this effect, a positive mass and a negative mass attract gravitationally while moving away from each other, thus violating the action-reaction principle. This effect has remained a major challenge for integrating negative masses into standard cosmology.\\

Thus, to resolve the crisis of modern cosmology, the Janus model proposes a paradigm shift. Since the 1970s, the $\Lambda$CDM model has failed to explain certain observed phenomena, such as the rotation speeds of galaxies and the acceleration of cosmic expansion. The Janus model, based on a bimetrical geometry with positive and negative masses evolving on distinct geodesics, offers an alternative. It proposes a new approach to solving anomalies such as the rapid formation of galaxies after the Big Bang and discrepancies in the measurement of the Hubble constant (\cite{Ferreira2022, Hoffman2017}).\\

The Janus model proposes a bimetrical system where gravitational interactions between positive and negative masses are described by distinct field equations, each associated with its own metric. The construction of a homogeneous, isotropic, and time-dependent solution in the Janus model relies on FLRW-type metrics, respectively describing the universes of positive and negative masses. A common energy conservation relation is established, proposing an exact solution for dust universes, where the observed cosmic acceleration is interpreted as a negative total energy. Numerical comparisons confirm the model's compatibility with observations, as illustrated by the magnitude-redshift curve. The interaction laws in the Janus model reveal that masses of the same sign attract, while those of opposite signs repel, thus eliminating the \textit{"runaway effect"}. The model reproduces local observations of general relativity while replacing dark matter and dark energy with invisible negative masses. These negative masses form void-like structures that confine positive mass, accelerating star and galaxy formation in the first few hundred million years, in agreement with data from the James Webb telescope and observations of large cosmic voids (\cite{DAgostini2018, Perlmutter1999, Petit1995, Petit2014, Riess1998}).\\

Finally, the mathematical consistency of the Janus model is demonstrated in the weak field limit, thanks to the generalized conservation of energy and the Bianchi identities. The calculation of Schwarzschild metrics for positive and negative masses shows that masses of the same sign attract, while those of opposite signs repel. The model satisfies the Tolman-Oppenheimer-Volkoff equations in the Newtonian approximation, while remaining compatible with cosmological observations. It is also valid in regions dominated by negative masses, such as the \textit{dipole repeller}, where it predicts a negative gravitational lensing effect, dimming the luminosity of background objects (\cite{Adler1975, Oppenheimer1939, Hoffman2017, aop2024,}).\\

In summary, our model proposes an extension of general relativity by introducing two distinct metrics, each associated with a type of mass, allowing for the explanation of both the acceleration of the universe's expansion and certain large-scale structures, while remaining compatible with local observations of general relativity. This analysis opens new perspectives and places the Janus model among the approaches that can be tested by modern cosmological observations.

\section{The physical interpretation of time Inversion (T-Symmetry)}

The T-symmetry refers to the inversion of the time coordinate. In 1970, contributing to the development of symplectic geometry and its application to physics, mathematician J.-M. Souriau provided the physical interpretation of this inversion of the time coordinate (\cite{Souriau1970structure}). The Gram matrix defining the \textit{Minkowski space} is :
\begin{equation}
	G=\begin{pmatrix}
		1 & 0 & 0 & 0 \\
		0 & -1& 0 & 0 \\
		0 & 0 & -1& 0 \\
		0 & 0 & 0 & -1
	\end{pmatrix}.
\end{equation}
Its isometry group is the \textit{Poincaré group}:
\begin{equation}
	\begin{pmatrix}
		L & C \\
		0 & 1
	\end{pmatrix}.
\end{equation}
Where $L$ is the matrix representing the \textit{Lorentz group} \( \mathcal{L}or \) which describes how spacetime coordinates change between different inertial frames. These transformations include rotations in space as well as Lorentz transformations (boosts), which are changes of reference frames moving at a constant speed relative to each other. it's axiomatically defined by:
\begin{equation}
	\transp{L}GL=G,
\end{equation}
and \( C \) is the quadrivector of space-time translations in \( \mathbb{R}^{1,3} \) as follows:
\begin{equation}
	C=\begin{pmatrix}
		\Delta t\\
		\Delta x\\
		\Delta y\\
		\Delta z
	\end{pmatrix}.
\end{equation}
It acts on points in \textit{Minkowski space}:
\begin{equation}
\xi = \begin{pmatrix}
    t \\
    x \\
    y \\
    z
\end{pmatrix}.
\label{chapter4_eq5}
\end{equation}

This Lie group with 10 independent parameters\footnote{Including the 6 independent parameters of the Lorentz group (3 rotations and 3 boosts) and 4 independent transformations, which are translations in the 4 directions of \textit{Minkowski space}.} is the isometry group of this space, defined by its metric:
\begin{equation}
\mathrm{d}s^2 = \mathrm{d}t^2 - \mathrm{d}x^2 - \mathrm{d}y^2 - \mathrm{d}z^2 .
\label{chapter4_eq6}
\end{equation}
The \textit{Lorentz group} \( \mathcal{L}or \) has four connected components:
\begin{itemize}
    \item \( \mathcal{L}or_n \) is the neutral component (its \textit{restricted subgroup}), does not invert either space or time and is defined by:
    $$\mathcal{L}or_n=\{L\in\mathcal{L}or,\ \det(L)=1\ \wedge\ [L]_{00}\geq 1\}$$
    \item \( \mathcal{L}or_s \) inverts space and is defined by:
    $$\mathcal{L}or_s=\{L\in\mathcal{L}or,\ \det(L)=-1\ \wedge\ [L]_{00}\geq 1\}$$
    \item \( \mathcal{L}or_t \) inverts time but not space and is defined by:
    $$\mathcal{L}or_t=\{L\in\mathcal{L}or,\ \det(L)=1\ \wedge\ [L]_{00}\leq -1\}$$
    \item \( \mathcal{L}or_{st} \) inverts both space and time and is defined by:
    $$\mathcal{L}or_{st}=\{L\in\mathcal{L}or,\ \det(L)=-1\ \wedge\ [L]_{00}\leq -1\}$$
\end{itemize}
And we have:
\begin{equation}
\mathcal{L}or = \mathcal{L}or_n \sqcup\ \mathcal{L}or_s \sqcup\ \mathcal{L}or_t \sqcup\ \mathcal{L}or_{st}.
\label{chapter4_eqLor}
\end{equation}
The first two components are grouped to form the so-called \emph{``orthochronous''} subgroup:
\begin{equation}
\mathcal{L}or_o = \mathcal{L}or_n \sqcup\ \mathcal{L}or_s.
\label{chapter4_eq7}
\end{equation}
It includes {P-symmetry}, which poses no problem for physicists who know that there are photons of \textit{"right"} and \textit{"left"} helicity whose motions are derived from this symmetry. This corresponds to the phenomenon of the polarization of light.\\

The last two components form the subset \emph{"retrochronous"} or \emph{"antichronous"}, whose components invert time:
\begin{equation}
\mathcal{L}or_a = \mathcal{L}or_t \sqcup\ \mathcal{L}or_{st}.
\label{chapter4_eq8}
\end{equation}
Thus, we have:
\begin{equation}
\mathcal{L}or = \mathcal{L}or_o \sqcup\ \mathcal{L}or_{a}.
\end{equation}
Noting that:
\begin{equation}
\mathcal{L}or_t = -\mathcal{L}or_s \quad \mathcal{L}or_{st} = -\mathcal{L}or_n.
\label{chapter4_eq9}
\end{equation}

The \textit{Poincaré group} inherits the properties of the \textit{Lorentz group} and thus has four connected components, it is defined by:
\begin{equation}
g := \left\{ \left( \begin{array}{cc}
L & C \\
0 & 1 \\
\end{array} \right),
\quad L \in \mathcal{L}or \land C \in \mathbb{R}^{1,3} \right\},
\label{chapter4_eqPoin}
\end{equation}
acting on \textit{Minkowski space} as follows:
\begin{equation}
g(X) = L.X + C . \\
\end{equation}

The action of the group on its space of moments is the action on the dual of the Lie algebra of the group\footnote{Souriau's approach, thanks to the Poincaré group which is the isometry group of \textit{Minkowski space} encompassing the \textit{Lorentz group} (with its four connected components), allows the parameters associated with each of these motions, whose representative points belong to a vector space, \textit{the space of moments}, to emerge. The dimension of this space is equal to that of the group: ten. Indeed, the \textit{Lorentz group} is made up of transformations that preserve the quadratic form of space-time. It consists of the orthochronous Lorentz transformations and the translation group. The transformations of the orthochronous \textit{Lorentz group} have 6 degrees of freedom, while the translation group has 4 degrees of freedom. This structure leads to 10 independent parameters of the Poincaré group. By combining them into an antisymmetrical matrix called a \textit{torsor}, the parameters of the space of motions can thus be defined.}. The element of the Lie algebra is obtained by differentiating the ten components of the group. Souriau designates by the Greek letter \(\Lambda\) the differential of the square matrix \(Z\) representing the element of the \textit{Poincaré group}, and by the Greek letter \(\Gamma\) the element of the subgroup of spatio-temporal translations\footnote{(13.54) of \cite{Souriau1970structure}. He then writes \(\mu\), an element of the space of motions, in the form (13.57) and expresses the invariance in the form of the constancy of the scalar (13.58), where M is an antisymmetric matrix.}:
\begin{equation}
Z := \left\{\begin{pmatrix} \Lambda & \Gamma \\ 0 & 0 \end{pmatrix}, \bar{\Lambda} = -\Lambda \land \Gamma \in \mathbb{R}^{1,3}\right\}.
\end{equation}

The elements of the \textit{Lorentz group} act on points in spacetime, transforming one motion into another. By applying an element \( L \) of the \textit{Lorentz group} to a given motion, we obtain a new motion.
%As mentioned through expression \ref{chapter4_eqLor}, the \textit{Lorentz group} has four connected components.\\
The neutral component \( \mathcal{L}or_n \) is a subgroup containing the identity matrix that inverts neither space nor time.\\

Let's consider the 4-component matrix \( \omega \) made up of two parameters \( \lambda_1 \) and \( \lambda_2 \):
\begin{equation}
\omega_{(\lambda_1, \lambda_2)} = \left( \begin{array}{cccc}
\lambda_1 & 0 & 0 & 0 \\
0 & \lambda_2 & 0 & 0 \\
0 & 0 & \lambda_2 & 0 \\
0 & 0 & 0 & \lambda_2
\end{array} \right) \quad \text{with} \quad \left\{
\begin{aligned}
\lambda_1 = \pm1 \\
\lambda_2 = \pm1
\end{aligned}
\right.
\end{equation}
Thus, the four components of the \textit{Lorentz group} can be easily expressed using the four possible combinations of these two parameters applied to its neutral component, of which an element \(L_n \in \mathcal{L}or_n\) is expressed according to the expression \(L = \omega L_n\):
\begin{equation}
\begin{array}{cc}
  \omega_{(1,1)} \times L_n = \left(
    \begin{array}{cccc}
      1 & 0 & 0 & 0 \\
      0 & 1 & 0 & 0 \\
      0 & 0 & 1 & 0 \\
      0 & 0 & 0 & 1
    \end{array}
  \right) \in \mathcal{L}or_n
  &
  \omega_{(1,-1)} \times L_n = \left(
    \begin{array}{cccc}
      1 & 0 & 0 & 0 \\
      0 & -1 & 0 & 0 \\
      0 & 0 & -1 & 0 \\
      0 & 0 & 0 & -1
    \end{array}
  \right) \in \mathcal{L}or_s
  \\
  \omega_{(-1,1)} \times L_n = \left(
    \begin{array}{cccc}
      -1 & 0 & 0 & 0 \\
      0 & 1 & 0 & 0 \\
      0 & 0 & 1 & 0 \\
      0 & 0 & 0 & 1
    \end{array}
  \right) \in \mathcal{L}or_t
  &
  \omega_{(-1,-1)} \times L_n = \left(
    \begin{array}{cccc}
      -1 & 0 & 0 & 0 \\
      0 & -1 & 0 & 0 \\
      0 & 0 & -1 & 0 \\
      0 & 0 & 0 & -1
    \end{array}
  \right) \in \mathcal{L}or_{st}
\end{array}
\end{equation}

We note that \( \lambda_1 = -1 \) inverts time while \( \lambda_2 = -1 \) inverts space. The four components are grouped into two subsets \emph{``orthochronous''} and \emph{``retrochronous''} according to the respective expressions \ref{chapter4_eq7} and \ref{chapter4_eq8}.\\

The \textit{Poincaré group} can then be written according to these four connected components as follows:
\begin{equation}
g := \left\{ \left( \begin{array}{cc}
\omega L_n & C \\
0 & 1
\end{array} \right),
\omega L_n \in \mathcal{L}or \land C \in \mathbb{R}^{1,3} \right\}.
\end{equation}
Thus, the action of this \textit{Poincaré group} on the spacetime coordinates yields the following space of motions:
\begin{equation}
\begin{bmatrix}
\omega L_n & C \\
0 & 1
\end{bmatrix}
\times
\begin{bmatrix}
\xi \\
1
\end{bmatrix}
=
\begin{bmatrix}
\omega L_n \xi + C \\
1
\end{bmatrix}.
\end{equation}
Indeed, this describes the action of the \textit{Poincaré group} on its space of moments \( \mu \), consisting of ten independent scalar quantities:
\begin{itemize}
  \item The energy \( E \),
  \item The momentum \( p = \{ p_x, p_y, p_z \} \),
  \item The passage \( f = \{ f_x, f_y, f_z \} \),
  \item The spin \( s = \{ l_x, l_y, l_z \} \).
\end{itemize}

If we consider the motion of an object in space. Such motion is also defined by its moment \(\mu\). The physicist can then apply an element \(\mathcal{G}\), for example from the Galilean group, to this moment \(\mu\). This produces a new moment \(\mu'\). This action can be written as follows:
\begin{equation}
\mu' = \mathcal{G}  \mu  \transp{\mathcal{G}},
\label{eq_groupe_1}
\end{equation}
where \(\transp{\mathcal{G}}\) represents the transpose of this matrix \(\mathcal{G}\). \(\mu\) is an antisymmetric moment matrix of size \(5 \times 5\)\footnote{Meaning the symmetric elements with respect to the main diagonal have opposite signs. The elements on the main diagonal are equal to zero, as each is its own opposite.} where the more compact form is defined as follows:
\begin{equation}
\mu = \left( \begin{matrix} M & -P \\ P^T & 0 \end{matrix} \right),
\label{eq_groupe2}
\end{equation}
with\footnote{\( M \) is the moment matrix associated with \(\mu\) with dimensions \( 4 \times 4 \), and \( P \), a four-vector energy-momentum with dimensions \( 4 \times 1 \).}:
\begin{equation}
M = \begin{pmatrix} 0 & -l_z & l_y & f_x \\ l_z & 0 & -l_x & f_y \\ -l_y & l_x & 0 & f_z \\ -f_x & -f_y & -f_z & 0 \end{pmatrix}, \quad P = \left( \begin{array}{c}
E \\
p_x \\
p_y \\
p_z 
\end{array} \right).
\label{eq_matricemoment}
\end{equation}

Then, by applying the action of the \textit{Poincaré group} \ref{chapter4_eqPoin} on the dual of its Lie algebra, i.e., on its space of moments, we obtain the following action according to \ref{eq_groupe_1} :
\begin{equation}
\mu' = \begin{pmatrix} L & C \\ 0 & 1 \end{pmatrix}  \begin{pmatrix} M & -P \\ \transp{P} & 0 \end{pmatrix}  \begin{pmatrix} \transp{L} & 0 \\ \transp{C} & 1 \end{pmatrix} ,
\end{equation}
\begin{equation}
\mu' = \begin{pmatrix} LM \transp{L} - LP \transp{C} + C \transp{P} \transp{L} & -LP \\ \transp{P} \transp{L} & 0 \end{pmatrix}.
\end{equation}
We can deduce\footnote{(13.107) of \cite{Souriau1970structure}.}:
\begin{equation}
	M'=LM\transp{L}+C\transp{P}\transp{L}-L\transp{P}C,
\end{equation}
and
\begin{equation}
	P'=LP.
	\label{eq_pprime}
\end{equation}
Therefore, the torsor of \textit{Poincaré group} is given by the different components of the space of moments\footnote{(13.57) of \cite{Souriau1970structure}.} as follows:
\begin{equation}
\mu = \{M,P\} = \{l,g,p,E\},
\end{equation}
where \(l\) is the angular momentum of \(M\), \(g\) is the relativistic barycenter of \(M\), \(p\) is the linear momentum of \(P\) and \(E\) is the energy of \(P\). \\

Now, let's consider for example the symmetry T, where there is only a time inversion (\( \lambda_1 = -1 \)), without space inversion (\( \lambda_2 = 1 \)), in a case where there is also no translation in spacetime (\( C = 0 \)). We thus have:
\begin{equation}
\omega_{(-1, 1)} \times L_n = L_t .
\end{equation}
Hence:
\begin{equation}
L_t \times \xi = \left( \begin{array}{cccc}
-1 & 0 & 0 & 0 \\
0 & 1 & 0 & 0 \\
0 & 0 & 1 & 0 \\
0 & 0 & 0 & 1
\end{array} \right) \times
\left( \begin{array}{c}
t \\
x \\
y \\
z 
\end{array} \right) =
\left( \begin{array}{c}
\textit{\textbf{-t}} \\
x \\
y \\
z
\end{array} \right).
\end{equation}
Thus, we obtain the action of time inversion in the space of trajectories or in spacetime. \\

The second equation (\ref{eq_pprime}) sheds light on the physical significance of this inversion of the time coordinate. Indeed, the application of the \( L_t \) component of the \textit{Lorentz group} to the motion of a particle gives:
\begin{equation}
P' = L_t P = 
\left( \begin{array}{cccc}
-1 & 0 & 0 & 0 \\
0 & 1 & 0 & 0 \\
0 & 0 & 1 & 0 \\
0 & 0 & 0 & 1
\end{array} \right)
\left( \begin{array}{c}
E \\
p_x \\
p_y \\
p_z 
\end{array} \right) =
\left( \begin{array}{c}
\textit{\textbf{-E}} \\
p_x \\
p_y \\
p_z 
\end{array} \right).
\end{equation}
Therefore, we can deduce that the application of the \( L_t \) component of the \textit{Lorentz group} to the motion of a particle induces an inversion of its energy from \( E \) to \( \textit{\textbf{-E}} \).\\% and its passage from \( f \) to \( \textit{\textbf{-f}} \). \\

The T symmetry applied to the motion of a particle inverts its energy which leads to mass inversion\footnote{page 198-199 of \cite{Souriau1970structure}.} following the definition of the mass\footnote{(14.57) on page of \cite{Souriau1970structure}.} as:
\begin{equation}
m = \sqrt{\transp{P} \cdot P} \, \sgn(E).
\end{equation}
A very detailed commentary on the work can be found in reference \cite{desaxce2023presentation}. The approach is based on the introduction of the space of motions as a dual of the Lie algebra of the group.\\ 

In this context, we uncover the physical interpretation of the model proposed by A. Sakharov: the second universe in his framework could consist of particles possessing both negative energy and negative mass.\\

To further extend the interpretation of fundamental symmetries, we now turn our attention to C-symmetry, which is associated with charge conjugation. By introducing a higher-dimensional framework inspired by Kaluza-Klein theory, we can offer a geometrical interpretation of electric charge, according to Noether's theorem. This will allow us to explore the relationship between spacetime transformations and the emergence of electrically charged particles.

\section{Geometrical interpretation of electric charge}

The geometrical interpretation of C-symmetry, which is synonymous with charge conjugation and matter-antimatter duality, was provided by J.-M. Souriau in 1964 in chapter V of reference \cite{Souriau1970structure}.\\

Let's apply an extension of the \textit{Poincaré group} to form the following dynamic group:
\begin{equation}
g := \left\{ \left( \begin{array}{ccc}
1 & 0 & \phi \\
0 & L & C \\
0 & 0 & 1 \end{array} \right),
\phi \in \mathbb{R} \land L = \lambda L_o \in \mathcal{L}or \land \lambda = \pm 1 \land C \in \mathbb{R}^{1,3} \right\}.
\end{equation}
Starting from \textit{Minkowski space}:
\begin{equation}
\xi = \begin{pmatrix} t \\ x \\ y \\ z \end{pmatrix} = \begin{pmatrix} t \\ r \end{pmatrix},
\end{equation}
let's introduce Kaluza space\footnote{Kaluza space is a hyperbolic Riemannian manifold with signature $(+ - - - -)$.} that incorporates a $5 \times 5$ Gram matrix:
\begin{equation}
\Gamma = \begin{pmatrix}
1 & 0 & 0 & 0 & 0 \\
0 & -1 & 0 & 0 & 0 \\
0 & 0 & -1 & 0 & 0 \\
0 & 0 & 0 & -1 & 0 \\
0 & 0 & 0 & 0 & -1 \\
\end{pmatrix}
= \begin{pmatrix}
G & 0 \\
0 & -1 \\
\end{pmatrix}
\quad \text{where} \quad G = \begin{pmatrix}
1 & 0 & 0 & 0 \\
0 & -1 & 0 & 0 \\
0 & 0 & -1 & 0 \\
0 & 0 & 0 & -1 \\
\end{pmatrix}.
\end{equation}

In the considered group, we just add a translation \(\phi\) to the fifth dimension \(\zeta\). Thus, the dimension of the group becomes 11. It is the isometry group of Kaluza space, defined by its metric:
\begin{equation}
\mathrm{d}s^2 = \transp{\mathrm{d}X} \Gamma {\mathrm{d}X} = \mathrm{d}t^2 - \mathrm{d}x^2 - \mathrm{d}y^2 - \mathrm{d}z^2 - \mathrm{d}\zeta^2,
\label{eq_ds2_kaluza1}
\end{equation}
with :
\begin{equation}
X = \begin{pmatrix} \xi \\ \zeta \end{pmatrix} = \begin{pmatrix} t \\ x \\ y \\ z \\ \zeta \end{pmatrix}.
\label{eq_groupe_k2}
\end{equation}
According to Noether's theorem\footnote{Noether's theorem states that for every continuous symmetry of a physical action, there exists a conserved quantity. In our context, if a new symmetry ensures the invariance of a scalar \(q\), this scalar is the conserved quantity. This means that \(q\) remains constant when the symmetry is applied to the system's action.}, this new symmetry is accompanied by the invariance of a scalar that we will call \(q\). The torsor of this \textit{Kaluza group} then incorporates an additional parameter:
\begin{equation}
\mu = \{M,P,q\} = \{l,g,p,E,q\}.
\label{eq_ds2_kaluza2}
\end{equation}

Let's introduce the action of the group on its Lie algebra:
\begin{equation}
Z' = g^{-1} Z g.
\label{eq_groupe_k1}
\end{equation}
If we consider an element of the Lie algebra of this group:
\begin{equation}
Z = \begin{pmatrix}
0 & 0 & \delta\phi \\
0 & G\omega & \gamma \\
0 & 0 & 0
\end{pmatrix}
\quad
Z' = \begin{pmatrix}
0 & 0 & \delta\phi' \\
0 & G\omega' & \gamma' \\
0 & 0 & 0
\end{pmatrix},
\end{equation}
we obtain: 
\begin{equation}
Z' =
\begin{pmatrix}
0 & 0 & \delta\phi' \\
0 & G\omega' & \gamma' \\
0 & 0 & 0
\end{pmatrix}
=
\left(
\begin{array}{ccc}
0 & 0 & \delta\phi \\
0 & L^{-1}G\omega L & L^{-1}G\omega C + L^{-1}\gamma \\
0 & 0 & 0 \\
\end{array}
\right).
\end{equation}
%Then, we can deduce :
%\begin{equation}
%\frac{1}{2} T_r (M \cdot \omega) + P^T \cdot G\gamma + q\delta\phi = \frac{1}{2} T_r (M' \cdot \omega') + {P'}^T \cdot G\gamma' + q'\delta\phi'
%\end{equation}
This allows us to deduce the action of the following group:
\begin{align}
q' &= q,\\
M' &= LM \transp{L} - LP \transp{C} + C \transp{P} \transp{L} ,\\
P' &= LP.
\end{align}
If we identify \(q\) as the electric charge, this would show that the motion of a massive particle in a five-dimensional space would transform it into an electrically charged particle.\\

The interpretation of C-symmetry within a higher-dimensional framework, as explored, leads naturally to a broader geometric understanding of symmetries in the Janus model. Specifically, the notion of charge conjugation extends to encompass the duality between matter and antimatter. To develop this further, we now introduce the \textit{Janus restricted group}, which provides a formal structure to describe these symmetries. This group allows us to explore how quantum charges can be inverted by compactified dimensions, linking the symmetry properties of spacetime to the emergence of quantized charges and new quantum numbers.

\section{Matter-antimatter symmetry (C-symmetry)}

Let's introduce the \textit{Janus restricted group} as follows:
\begin{equation}
g := \left\{ \left( \begin{array}{ccc}
\mu & 0 & \phi \\
0 & L & C \\
0 & 0 & 1 \end{array} \right),
\mu = \pm 1 \land \phi \in \mathbb{R} \land L = \lambda L_o \in \mathcal{L}or \land \lambda = \pm 1 \land C \in \mathbb{R}^{1,3} \right\}.
\end{equation}
The action of the group on the coordinates of the 5-dimensional spacetime defined by \ref{eq_groupe_k2} yields the space of the following motions:
\begin{equation}
\left(
\begin{array}{ccc}
\mu & 0 & \phi \\
0 & L & C \\
0 & 0 & 1 \\
\end{array}
\right)
\left(
\begin{array}{c}
\zeta \\
\xi \\
1 \\
\end{array}
\right)
=
\left(
\begin{array}{c}
\mu \zeta + \phi \\
L \xi + C \\
1 \\
\end{array}
\right).
\end{equation}
A similar calculation to the previous one yields the action of the group:
\begin{align}
q' &= \mu q,\\
M' &= LM \transp{L} - LP \transp{C} + C \transp{P} \transp{L}, \\
P' &= LP.
\end{align}
This group acts on the five-dimensional Kaluza space. We observe that \( \mu = -1 \) reverses the fifth dimension \(\zeta\) and the scalar \(q\).\\

Through a dynamic interpretation of the group, we find the idea suggested by J.-M. Souriau \cite{Souriau1970structure}: the inversion of the fifth dimension is associated with the inversion of electric charge. However, this is only one of the quantum charges. 
Indeed, the {C-Symmetry} translating the \textit{"matter-antimatter"} symmetry introduced by Dirac (\cite{dirac1928quantum}), reverses all quantum charges. This inversion operation is only obtained by adding as many compactified dimensions as there are quantum charges. The action of the group on the coordinates of \(n\)-dimensional spacetime yields the space of the following motions:
\begin{equation}
\left(
\begin{array}{cccccc}
\mu & 0 & 0 & \cdots & 0 & \phi^1 \\
0 & \mu & 0 & \cdots & 0 & \phi^2 \\
0 & 0 & \ddots & \cdots & 0 & \vdots \\
\vdots & \vdots & \cdots & \mu & 0 & \phi^p \\
0 & 0 & \cdots & 0 & L & C \\
0 & 0 & \cdots & 0 & 0 & 1 \\
\end{array}
\right)
\left(
\begin{array}{c}
\zeta^1 \\
\zeta^2 \\
\vdots \\
\zeta^p \\
\xi \\
1 \\
\end{array}
\right)
=
\left(
\begin{array}{c}
\mu \zeta^1 + \phi^1 \\
\mu \zeta^2 + \phi^2 \\
\vdots \\
\mu \zeta^p + \phi^p \\
L \xi + C \\
1 \\
\end{array}
\right).
\end{equation}
The torsor of this group incorporates several additional scalars $q^p$:
\begin{equation}
\mu = \{M,P,\sum_{1}^{p} q^i\} = \{l,g,p,E,q^1,q^2,\hdots,q^p\}.
\end{equation}
This allows us to obtain the action of the group on its momentum space:
\begin{align}
{q'}^1 &= \mu q^1,\\
{q'}^1 &= \mu q^1,\\
\hdots \\
{q'}^p &= \mu q^p,\\
M' &= LM \transp{L} - LP \transp{C} + C \transp{P} \transp{L}, \\
P' &= LP.
\end{align}

Moreover, Souriau considers that electric charge can be geometrically quantized into discrete values \( (+e, 0, -e) \) when the associated fifth dimension is closed.\\

\begin{figure}[h!]
\begin{center}
\includegraphics[width=\textwidth]{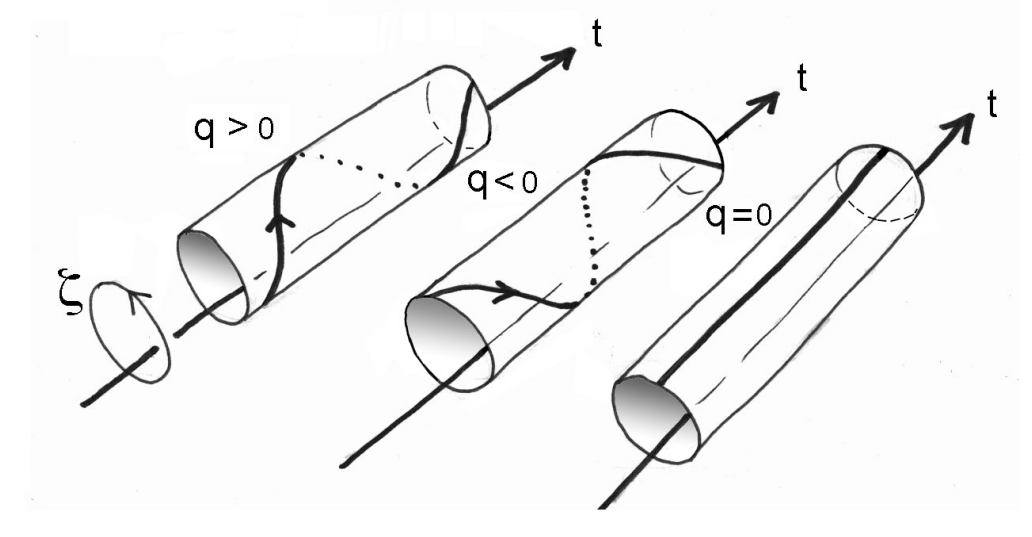}
\end{center}
\caption{Inversion of the winding direction of a particle's motion reflecting the C-symmetry}
\label{fig:C-symetrie}
\end{figure}

Imagine representing motion in \textit{Minkowski space} along a simple straight line oriented in time. At each point, we add a closed dimension, which extends \textit{Minkowski space} into a bundle. In the didactic \ref{fig:C-symetrie}, it is represented as a cylinder.

But in approach \cite{desaxce2024symmetry}, these transformations no longer a priori preserve the electric charge $q$, which then becomes dependent on the chosen coordinate system. In reference \cite{desaxce2024symmetry}, taking up the approach initiated in \cite{SouriauGeomRelat}, the author opts for a closed fifth dimension, in which the radius of this \textit{"universe tube"} becomes very small, of the order of Planck's length. He then rediscovers the invariance of electric charge and concludes \cite{desaxce2024symmetry}, we quote:

\begin{quotation}
In this paper, we revisit the Kaluza-Klein theory from the perspective of the classification
of elementary particles based on the coadjoint orbit method. The keystone conjecture is to
consider the electric charge as an extra momentum on an equal footing with the mass and the
linear momentum. We study the momentum map of the corresponding symmetry group $\hat{\mathbb{G}}_1$
which conserves the hyperbolic metric. We show that the electric charge is not an invariant,
\emph{i.e.} it depends on the reference frame, which is in contradiction with the experimental
observations. In other words, it is not the symmetry group of the Universe today as we
know it. To avert this paradox, we scale the fifth coordinate and consider the limit when the
cylinder radius $\omega$ vanishes. For the corresponding group $\hat{\mathbb{G}}_0$ also of dimension 15, the charge
is an invariant then independent of the frame of reference and the observer. On this ground,
we propose a cosmological scenario in which the elementary particles of the early Universe
are classified from the momenta of the group $\hat{\mathbb{G}}_1$, next the three former dimensions inflate
quickly while the fifth one shrinks, leading to the 4D era in which as today the particles are
characterized by the momenta of the group $\hat{\mathbb{G}}_0$. By this mechanism, the elementary particles
can acquire electric charge as a by-product of the $4 + 1$ symmetry breaking of the Universe.
This work opens the way to the geometric quantization of charged elementary particles.
\end{quotation}
The expression for this characteristic dimension of this universal tube is given in \cite{SouriauGeomRelat} on page 412 :

\begin{equation}
	\mathit{e}\frac{\hbar}{e}\sqrt{\frac{\chi}{2\pi}}, % ici, on laisse e en italique, c'est la charge électrique élémentaire, pas la base des logarithmes naturels.
\end{equation}
$\chi$ being the Einstein constant taken equal to \cite{SouriauGeomRelat}:

\begin{equation}
	\chi=-\frac{8\pi G}{c^2}=\qty{1.856e-27}{\centi\meter\per\gram}.
\end{equation}
By introducing numerical values, this characteristic length is \qty{3.782e-32}{\centi\meter}. Dividing by $2\pi$ gives us the order of magnitude of Planck's length. In this view, the quantization of electric charge and its constancy are derived from the closure of the extra dimension associated with the decrease in the characteristic dimension associated with it.\\

This group refers to an extension of the \textit{Poincaré group}, i.e.~to a field-free, curvature-free universe. This construction of a five-dimensional relativity was suggested in 1964 in reference \cite{SouriauGeomRelat} and has been taken up again more recently in \cite{desaxce2024symmetry}. Note that it is in \cite{SouriauGeomRelat}, page 413, that the link between charge conjugation and fifth-dimensional inversion is first mentioned.\\

By generalizing \cite{Kerner}, we can envisage an extension of space-time to a space with $4 + p$ dimensions, all of which may see their characteristic dimensions reduced, like that of this fifth dimension, each of these collapses leading to the emergence and quantization of new quantum numbers, baryonic, leptonic, unique etc., the electric charge being only the first of these.\\

Thus, the \textit{Janus restricted group} has provided us with a framework for understanding the matter-antimatter symmetry (C-symmetry) and the inversion of quantum charges through additional compactified dimensions. We can now extend it to a broader symmetry group associated with A. Sakharov's model, the \textit{Janus group}, which incorporates both C-symmetry and PT-symmetry. This extension allows us to explore a dynamic group structure that includes negative masses and antimatter within the framework of Sakharov's twin universe model.

\section{Group associated with A. Sakharov's model: the \textit{Janus group}}

If we want to construct a group that translates the T-symmetry invoked by Sakharov, we'll replace $\Loo$ by $\lambda\Loo$ with $\lambda = \pm 1$. But, as proposed in \cite{Petit2014}, we can translate what had already been proposed \cite{Sakharov1967}, we quote:

\begin{quotation}
	All phenomena corresponding to $t <0$ are, in this hypothesis, assumed to be CPT images of phenomena corresponding to $t > 0$.
\end{quotation}
Then, by introducing a new symmetry to the previous \textit{Janus restricted group}, which we can call {PT Symmetry} allowing the conversion of matter into antimatter with negative mass\footnote{A concept we could call \textit{antimatter in the sense of Feynman} (\cite{feynman1949}).}, we thus combine C-symmetry and PT-symmetry to form the \textit{Janus dynamic group} as follows: 
\begin{equation}
g := \left\{ \left( \begin{array}{ccc}
\lambda\mu & 0 & \phi \\
0 & \lambda L_o & C \\
0 & 0 & 1 \end{array} \right),
\lambda, \mu \in \{-1,1\} \land \phi \in \mathbb{R} \land L_o \in \mathcal{L}or_o \land C \in \mathbb{R}^{1,3} \right\}.
\label{eq_gjanus}
\end{equation}

We can consider that particles of matter and antimatter can coexist in the same space fold. However, no coexistence is possible for the motion of particles deduced by T-symmetry (or PT-symmetry). This space is of dimension \(4 + p\) (for \(p\) quantum charges). We will therefore consider the two-fold covering of this manifold \(M_{n+p}\). In each of these two folds, there remains a possibility to perform the symmetry corresponding to \(\mu = -1\), that is, the inversion of all quantum charges. In other words, the \textit{"matter-antimatter"} duality exists in both folds.\\

To understand the nature of the different components of these folds, we will consider the motion of a particle of matter with energy and mass:

\begin{itemize}
  \item By acting on this motion with elements of the group corresponding to \((\lambda = 1; \mu = 1)\), we will obtain other motions of particles of matter with positive mass and energy.
  
  \item By acting on this motion with elements of the group corresponding to \((\lambda = 1; \mu = -1)\), we will obtain other motions of antimatter particles with positive mass and energy\footnote{These are \textit{"antimatter in the sense of Dirac"} (C-symmetry).}.
  
  \item By acting on this motion with elements of the group corresponding to \((\lambda = -1; \mu = 1)\), we will obtain other motions of particles of matter with negative mass and energy\footnote{CPT-symmetry.}.
  
  \item By acting on this motion with elements of the group corresponding to \((\lambda = -1; \mu = -1)\), we will obtain other motions of antimatter particles with negative mass and energy\footnote{These are \textit{"antimatter in the sense of Feynman"} (PT-symmetry).}.
\end{itemize}

Its isometry group is that of Janus space, defined by the same metric as structuring Kaluza space (\ref{eq_ds2_kaluza1}), and its dimension is 11\footnote{10 + 1 dimension associated with the fifth space dimension \(\zeta\) that J.-M. Souriau identifies with the electric charge \(q\).}. The torsor of the group is also the same as (\ref{eq_ds2_kaluza2}).\\
However, if we consider an element of the Lie algebra of this group:
\begin{equation}
Z =
\begin{pmatrix}
0 & 0 & \delta\phi \\
0 & \lambda G \omega & \gamma \\
0 & 0 & 1
\end{pmatrix},
\end{equation}
we can then calculate \(Z'\) according to the relation \ref{eq_groupe_k1} as follows:
\begin{equation}
Z' = \begin{pmatrix}
0 & 0 & \delta\phi' \\
0 & \lambda G \omega' & \gamma' \\
0 & 0 & 1
\end{pmatrix}
= \left(
\begin{array}{ccc}
0 & 0 & (\lambda\mu) \delta\phi \\
0 & \lambda^3 L_o^{-1} G\omega L_o & \lambda^2 L_o^{-1} G\omega C + \lambda L_o^{-1} \gamma \\
0 & 0 & 0
\end{array}
\right).
\end{equation}
Thus, by identification, we can deduce:
\begin{align}
\delta\phi' &= \lambda\mu\delta\phi ,\\
\omega' &= \lambda^2 GL_o^{-1} G\omega L_o ,\\
\gamma' &= \lambda^2 L_o^{-1} G\omega C + \lambda L_o^{-1} \gamma .
\end{align}
We know that:
\begin{align}
L_o^{-1} &= G \transp{L_o} G.
\end{align}
Then: 
\begin{equation}
\begin{aligned}
\delta\phi' &= \lambda\mu\delta\phi ,\\
\omega' &= \lambda^2 \transp{L_o} \omega L_o ,\\
\gamma' &= \lambda^2 G \transp{L_o} \omega C + \lambda G \transp{L_o} G\gamma .
\end{aligned}
\label{eq:groupe_janus}
\end{equation}
However, inspired by J.-M. Souriau, we could add as many additional closed dimensions as quantum charges and write the dynamic group as follows:
\begin{equation}
\left(
\begin{array}{cccccc}
\lambda\mu & 0 & 0 & \cdots & 0 & \phi^1 \\
0 & \lambda\mu & 0 & \cdots & 0 & \phi^2 \\
0 & 0 & \ddots & \cdots & 0 & \vdots \\
\vdots & \vdots & \cdots & \lambda\mu & 0 & \phi^p \\
0 & 0 & \cdots & 0 & \lambda L_o & C \\
0 & 0 & \cdots & 0 & 0 & 1 \\
\end{array}
\right).
\end{equation}

The isometry group of this space can be defined by the following metric:
\begin{equation}
\mathrm{d}s^2 = (\mathrm{d}t)^2 - (\mathrm{d}x)^2 - (\mathrm{d}y)^2 - (\mathrm{d}z)^2 - (\mathrm{d}{\zeta^1})^2 - (\mathrm{d}{\zeta^2})^2 - \hdots - (\mathrm{d}{\zeta^p})^2 .
\end{equation}
With :
\begin{equation}
X = \begin{pmatrix} \xi \\ \zeta \end{pmatrix} = \begin{pmatrix} t \\ x \\ y \\ z \\ \zeta^1 \\ \zeta^2 \\ \vdots \\ \zeta^p \end{pmatrix}.
\end{equation}
The action of this \textit{Janus group} on the coordinates of \(10+p\) independant parameters then yields the space of the following motions:
\begin{equation}
\left(
\begin{array}{cccccc}
\lambda\mu & 0 & 0 & \cdots & 0 & \phi^1 \\
0 & \lambda\mu & 0 & \cdots & 0 & \phi^2 \\
0 & 0 & \ddots & \cdots & 0 & \vdots \\
\vdots & \vdots & \cdots & \lambda\mu & 0 & \phi^p \\
0 & 0 & \cdots & 0 & \lambda L_o & C \\
0 & 0 & \cdots & 0 & 0 & 1 \\
\end{array}
\right)
\left(
\begin{array}{c}
\zeta^1 \\
\zeta^2 \\
\vdots \\
\zeta^p \\
\xi \\
1 \\
\end{array}
\right)
=
\left(
\begin{array}{c}
\lambda\mu \zeta^1 + \phi^1 \\
\lambda\mu \zeta^2 + \phi^2 \\
\vdots \\
\lambda\mu \zeta^p + \phi^p \\
\lambda L_o \xi + C \\
1 \\
\end{array}
\right).
\end{equation}

According to Noether's theorem, this new symmetry is accompanied by the invariance of additional scalars \(q^p\). Therefore, the torsor of the group integrates them according to this relation:
\begin{equation}
\mu = \{M,P,\sum_{1}^{p} q^i\} = \{l,g,p,E,q^1,q^2,\hdots,q^p\}.
\end{equation}
Thus, the duality relation\footnote{(13.58) from \cite{Souriau1970structure}.} gives us:
\begin{equation}
\frac{1}{2} T_r (M \cdot \omega) + \transp{P} \cdot G\gamma + \delta\phi \sum_{1}^{p} q^i = \frac{1}{2} T_r (M' \cdot \omega') + \transp{P'} \cdot G\gamma' +  \delta\phi \sum_{1}^{p} {q'}^i.
\end{equation}
This allows us to deduce the action of the group by identification with \ref{eq:groupe_janus}:
\begin{align}
\sum_{1}^{p} {q'}^i &= \lambda\mu \sum_{1}^{p} q^i,\\
M' &= LM \transp{L} - LP \transp{C} + C \transp{P} \transp{L}, \\
P' &= LP.
\end{align}

Having established the \textit{Janus dynamic group} as a natural extension of Sakharov's model, which incorporates both PT-symmetry and C-symmetry, we now shift our focus to the topological implications of the Janus model. In particular, we will explore how the symmetries discussed earlier can emerge from a closed, higher-dimensional universe. This section delves into the topological structure of the model, illustrating how P and T symmetries can arise naturally from the geometry of a closed universe, modeled as a projective space $\mathbb{P}^4$.

\section{Topology of the Janus model}

\begin{figure}
	\centering
	\includegraphics[width=11cm]{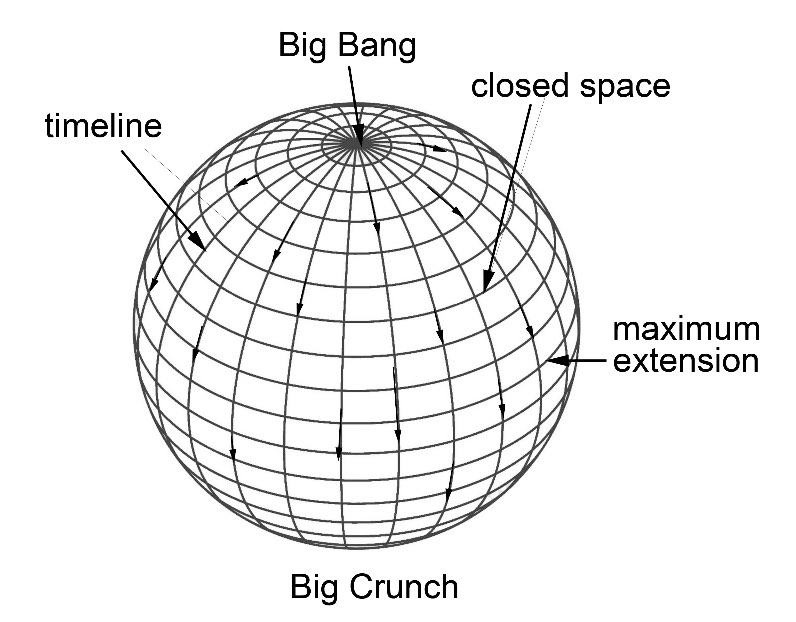}
	\caption{A simplified 2D representation of a closed universe with a spherical topology \( \mathbb{S}^2\), illustrating the temporal progression from the Big Bang to the Big Crunch, with the universe reaching maximum spatial extension in between.}\label{closed-universe}
\end{figure}

Let's consider a universe closed in all its dimensions, including space and time (see \ref{closed-universe}).
Diametrically opposed, antipodal points can be brought into coincidence. The image is then that of a $\mathbb{P}^2$ projective. The north and south poles, one representing the Big Bang and the other the Big Crunch, come into coincidence. The sphere cannot be paved without the presence of these two singularities. The same applies to any sphere $\mathbb{S}^{2n}$ if $n$ is even, especially if this dimension is 4. This geometry was proposed in \cite{Petit1994}.\\

\begin{figure}
	\centering
	\includegraphics[width=11cm]{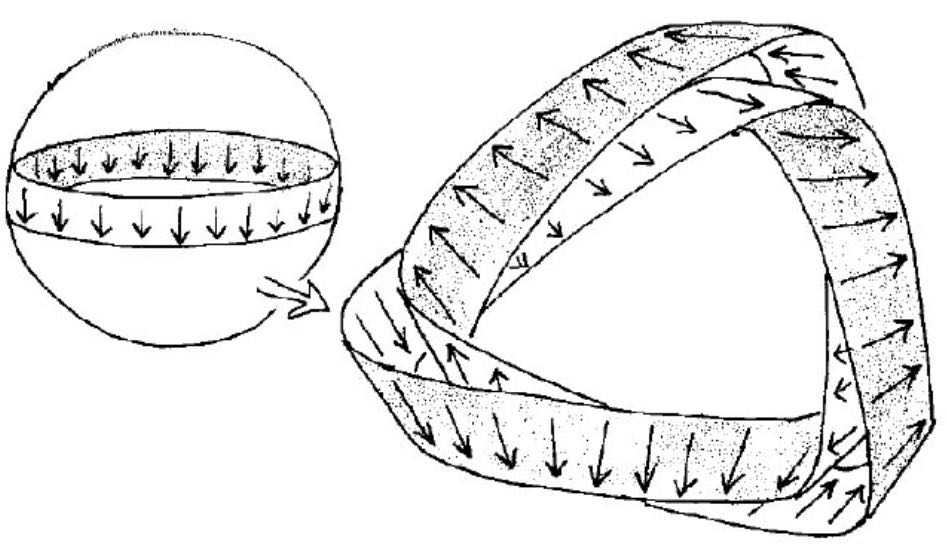}
	\caption{How the coincidence of antipodal regions creates T-symmetry. Drawing extracted from \cite{Petit-Topologicon}, page 65.}\label{antipodal-T-symmetry}
\end{figure}

\begin{figure}
	\centering
	\includegraphics[width=11cm]{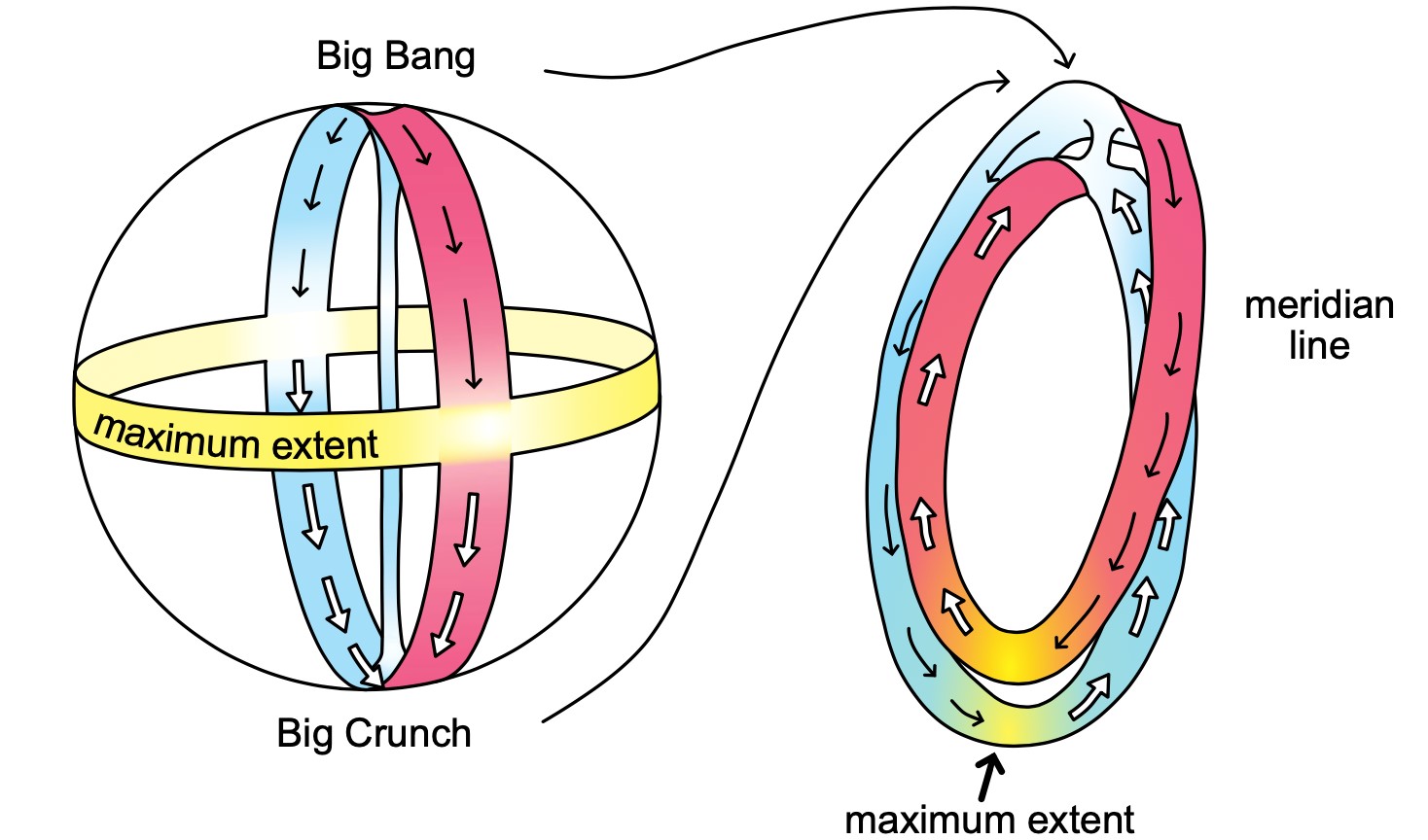}
	\caption{Coincidence of antipodal regions on a sphere $\mathbb{S}^2$, according to the two-folds cover of a half-turn Möbius strip, with the appearance of T-symmetry.}\label{antipodal-T-symmetry-Moebius}
\end{figure}

The \ref{antipodal-T-symmetry} shows how this coincidence of antipodal regions generates this T-symmetry. On the $\mathbb{S}^2$ sphere, the direction of time is given by the orientation of the meridian curves. This orientation is shown on the left at the new state of maximum expansion, when space is identified with the sphere's equator. During this folding of the $\mathbb{S}^2$ sphere, described in reference \cite{Petit-Topologicon} page 65, the vicinity of this equator is configured as the two-folds cover of a Möbius strip with three half-turns (see \ref{antipodal-T-symmetry} on the right).\\
\newpage
%\newpage
In \ref{antipodal-T-symmetry-Moebius}, we evoke the appearance of T-symmetry by manipulating the vicinity of a meridian line. In addition, we evoke the possible elimination of the Big Bang - Big Crunch double singularity by replacing them with a tubular passage, which then gives this geometry the nature of the two-fold cover of a Klein bottle.\\

\begin{figure}
	\centering
	\includegraphics[width=11cm]{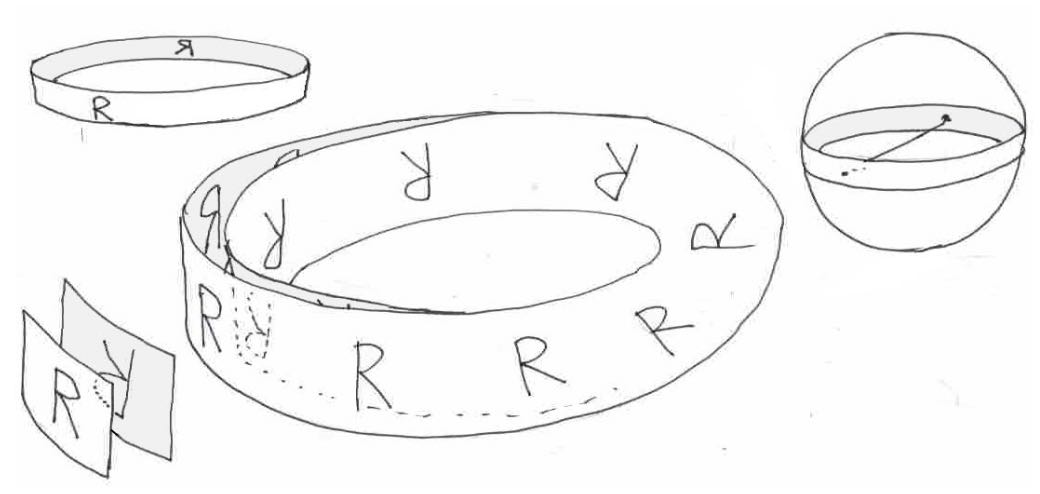}
	\caption{P-symmetry as a consequence of contacting antipodal region neighborhoods on an $\mathbb{S}^2$ sphere.}\label{P-symmetry}
\end{figure}

For enantiomorphy and P-symmetry to appear, the operation would have to be performed on a larger sphere. This aspect can be highlighted by considering the conjunction of antipodal regions in the vicinity of a meridian line, which is then configured according to the two-fold covering of a half-turn Möbius strip. The \ref{P-symmetry} illustrates this enantiomorphic situation.\\

\begin{figure}
	\centering
	\includegraphics[width=11cm]{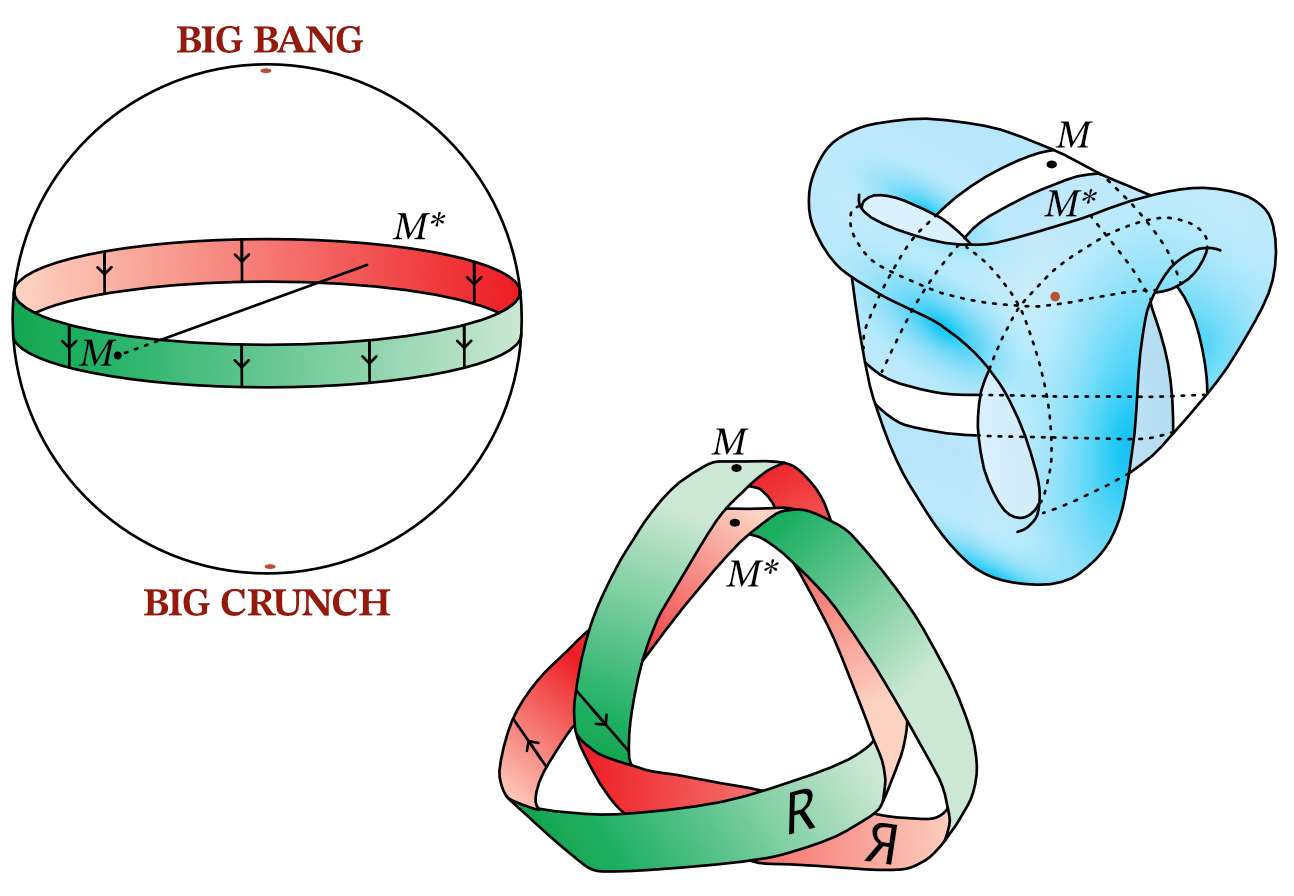}
	\caption{Boy's surface, immersion of the $\mathbb{P}^2$ projective in $\mathbb{R}^3$.}\label{Boy-surface}
\end{figure}

By bringing the antipodal points of even-dimensional spheres into coincidence, we locally create a configuration associating two T-symmetrical folds. By adding further dimensions, the coincidence of the antipodes creates a two-fold CPT-symmetric coating configuration of a projective space. In the case of the sphere $\mathbb{S}^2$, which corresponds only to a 2D didactic image, the image of the projective $\mathbb{P}^2$ is its immersion in, which corresponds to the surface described in 1903 by the German mathematician Werner Boy \cite{Boy1903}. See \ref{Boy-surface}. In this figure, we show how the coincidence of the antipodal points of the equator of the sphere $\mathbb{S}^2$ gives the two-fold covering of a Möbius ribbon with three half-turns.\\

In this section, we have demonstrated that the P and T symmetries invoked by A. Sakharov can arise as consequences of a purely topological structure, specifically the covering of a projective space $\mathbb{P}^4$.\\

After exploring the topological structure of the Janus model, we now address a major consequence of T-symmetry: the introduction of negative masses. According to J.-M. Souriau, the application of T-symmetry to the motion of a particle inverts its energy, which leads to the inversion of its mass\footnote{pages 198-199 of \cite{Souriau1970structure}.}, in accordance with the definition of mass\footnote{(14.57) on page 346 of \cite{Souriau1970structure}.}. Although this idea is elegant, it presents significant challenges when integrated into the framework of general relativity. In the following section, we will propose an initial approach to incorporating negative masses into the cosmological model, analyzing the implications of their interaction with positive masses and the resulting geodesics.

\section{Introducing negative masses: First approach}

Using dynamical group theory, we showed that this T-symmetry was synonymous with the introduction of negative masses into the cosmological model. A. Sakharov's primordial antimatter would therefore be endowed with negative mass. This first step is far from anecdotal since, if we neglect it, we admit to losing nothing less than half the universe from the outset. Is it then possible to introduce negative masses into the standard model of general relativity?\\

A first idea would be to consider that the field comes from two sources, represented by two tensors, the first referring to a positive mass content and the second to a negative mass content:
\begin{equation}
	\Rmunu-\frac{1}{2}R\gmunu=\chi\left[\Tmunu^\plus+\Tmunu^\moins\right].
\end{equation}
We can then consider the metric solution corresponding to a region where the field is created, firstly by a positive mass content:
\begin{equation}
	\Rmunu-\frac{1}{2}R\gmunu=\chi\Tmunu^\plus.
\end{equation}

\begin{figure}
	\centering
	\includegraphics[width=11cm]{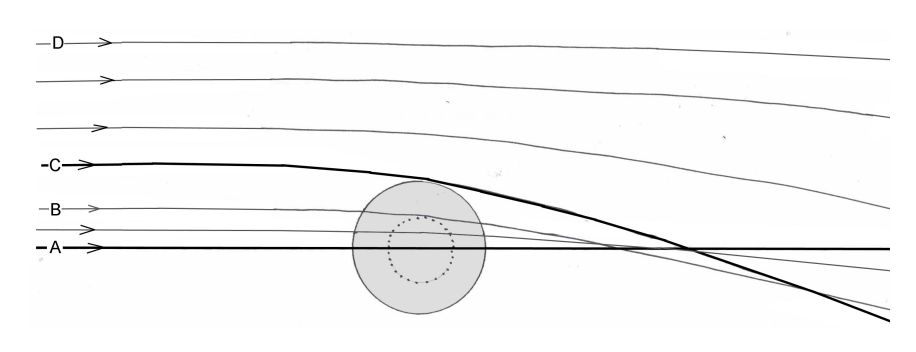}
	\caption{Deflection of positive-energy neutrinos by a positive mass. The trajectories, when passing near the mass, are deflected more strongly due to the gravitational effect. The angle of deflection reaches its maximum (C) when the neutrinos graze the edge of the mass. Trajectories further away, such as (D), experience a weaker deflection, and the deflection angle becomes null for trajectories passing at a very large distance from the mass. The trajectory passing through the center of the mass (A) remains undeflected due to the symmetry of the configuration.}
\label{geodesic-positive}
\end{figure}

Geodesics are given by a solution in the form of an external metric: 
\begin{equation}
	\di s^2=\left(1-\frac{2GM^\plus}{c^2r}\right)c^2\di t^2-\frac{\di r^2}{1-\dfrac{2GM^\plus}{c^2r}}-r^2\left(\di\theta^2+\sin^2\theta\di\varphi^2\right).
\end{equation}
The geodesics evoke an attraction (see \ref{geodesic-positive}).

\begin{figure}
	\centering
	\includegraphics[width=11cm]{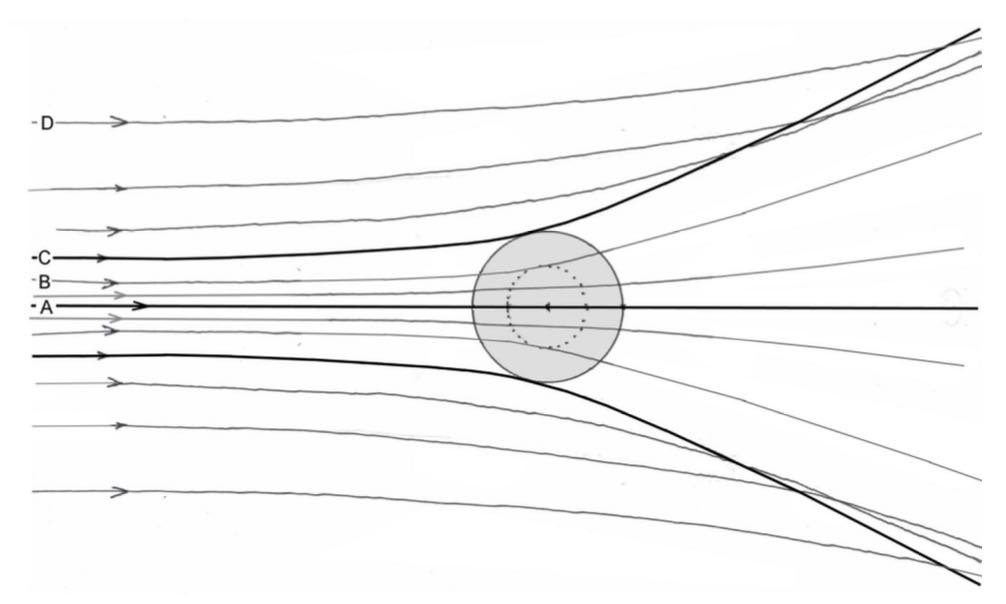}
	\caption{Deviation of positive-energy photons by a negative mass. The trajectories, when the curvature remains moderate, are very close to hyperbolas. The angle of deviation reaches a maximum (C) when the geodesic is tangent to the limit of the mass. It then decreases steadily to zero at very large distances (D). The angle of deviation is null, due to symmetry, when the geodesic passes through the center of the mass (A).}\label{photon-negative-mass}
\end{figure}

Now consider the field created by a negative mass $M^\moins$, the field equation becomes then:
\begin{equation}
	\Rmunu-\frac{1}{2}R\gmunu=\chi\Tmunu^\moins.
\end{equation}
And the solution corresponds to the metric:
\begin{equation}% utilisation de \abs de mismath pour un meilleur affichage des barres de valeur absolue. non, c'est trop moche, les barres sont gigantesques.
	\di s^2=\left(1+\frac{2G\abs{M^\moins}}{c^2r}\right)c^2\di t^2-\frac{\di r^2}{1+\dfrac{2G\abs{M^\moins}}{c^2r}}-r^2\left(\di\theta^2+\sin^2\theta\di\varphi^2\right).
\end{equation}
The geodesics then represent a repulsion (see \ref{photon-negative-mass}).\\

%\begin{figure}
%	\centering
%	\includegraphics[width=11cm]{images/geodesic-negative}
%	\caption{Geodesics created by negative mass.}\label{geodesic-negative}
%\end{figure}

In this context, our single field equation provides only a single family of geodesics, which the test particles, with both positive and negative masses, must follow. We deduce that:

\begin{itemize}
\item Positive masses attract both positive and negative masses.

\item Negative masses repel both positive and negative masses.

\item Two masses of identical absolute values but opposite signs are brought together, the positive mass flees, pursued by the negative mass. Both then accelerate uniformly, but without any energy input, since the energy of the negative mass is itself negative. This result was illustrated in 1957 by H. Bondi \cite{Bondi}. This phenomenon is known as \textit{"runaway effect"}. What's more, this scheme violates the action-reaction principle. In 1957, the conclusion was reached that it was physically impossible to include negative masses in the cosmological model. This would only be possible at the price of a profound paradigmatic shift, not by denying the achievements of general relativity, but by considering its extension in a wider geometric context.
\end{itemize}

After examining the introduction of negative masses and their implications within the framework of general relativity, we now turn to a broader cosmological context. The discovery of anomalies, such as the dipole repeller and the accelerating expansion of the universe, has revealed significant shortcomings in the standard model $\Lambda$CDM. Recent observations, particularly those made with the James Webb Space Telescope, have intensified the crisis in cosmology by challenging long-held assumptions about galaxy formation. In the following section, we will explore how the Janus cosmological model offers a paradigm shift capable of resolving these issues by proposing a bimetrical structure for the universe, integrating both positive and negative masses into a broader and more innovative geometric framework.

\section{A paradigm shift to escape the crisis of today's cosmology}

In the mid-1970s, the excessive rotation speeds of stars in galaxies had already led specialists to propose the existence of dark matter, ensuring their cohesion. In 2011, the discovery that the cosmic expansion was accelerating was attributed to a new, unknown ingredient known as dark energy.{\vspace{-38em}\marginparwidth1.5in \vspace{38em}} Over the decades, all attempts to assign an identity to these new components ended in failure.\\

\begin{figure}
	\centering
	\includegraphics[width=14cm]{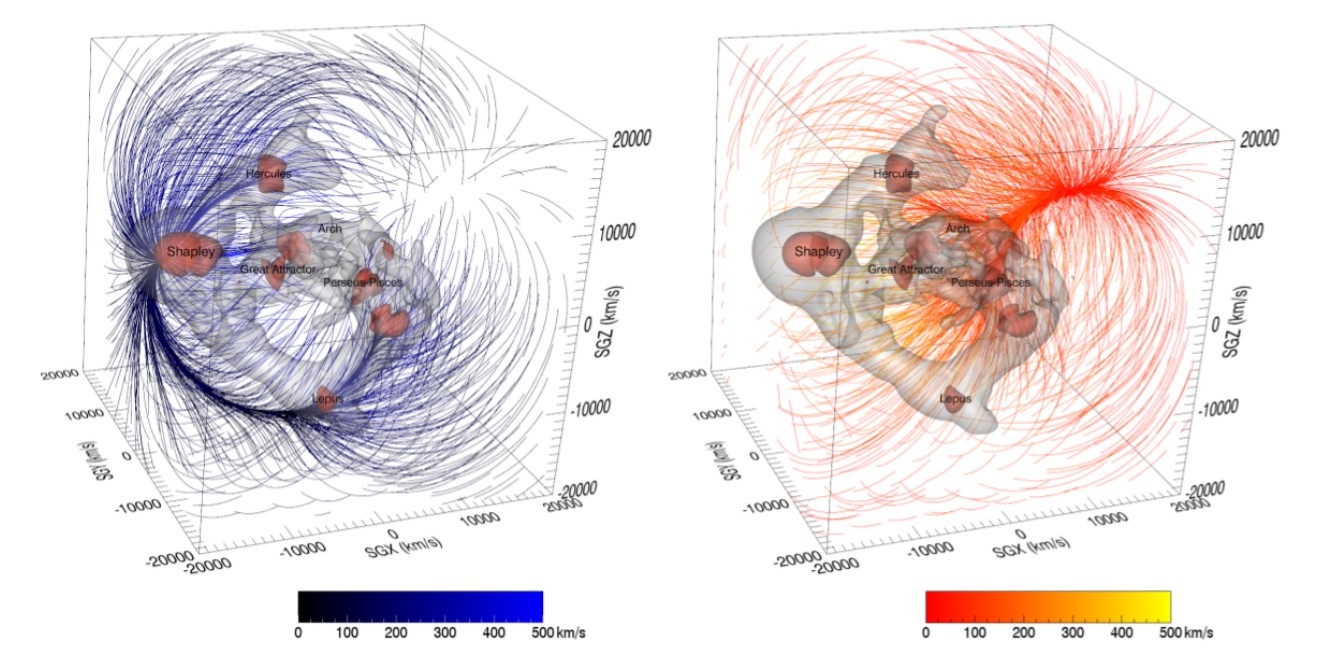}
	\caption{Location of the dipole repeller (highlighted by the red circle) within the large-scale structure of the universe~\cite{Hoffman2017}. The dipole repeller is a hypothesized region of space where galaxies are pushed away from, counteracting the attractive force of the Shapley Supercluster.}\label{dipole-repeller}
\end{figure}

In 2017 \cite{Hoffman2017}, Hélène Courtois, Daniel Pomarède, Brent Tully and Yeudi Hoffman produced the first very-large-scale mapping of the universe, in a cube of one and a half billion light-years across, with the Milky Way, our observation point, at the center (see \vref{dipole-repeller}). By subtracting the radial component of the velocity linked to the expansion motion, they indicate the trajectories followed by the masses. A dipolar structure appears. One formation, the Shapley attractor, comprising hundreds of thousands of galaxies, attracts galaxies to itself. But, symmetrically to this formation, 600~million light-years from the Milky Way, there is an immense void, some one hundred million light-years across, which, on the contrary, repels galaxies, and to which we give the name of dipole repeller. To date, no theory has been able to explain the existence of this vast void. While the idea of a gap in dark matter, positive and attractive, has been evoked, it doesn't hold water, as no mechanism has been found to give rise to it. Since 2017, several other such voids have been detected and located.\\

The launch of the James Webb Space Telescope has only added to this crisis \cite{Ferreira2022}. The Standard Model $\Lambda$CDM proposes a hierarchical mechanism for the birth of stars and galaxies. Gravitational instability appears as soon as matter and radiation are decoupled. The scenarios for the formation of both stars and galaxies in this model make use of the attributes conferred on hypothetical dark matter. But even with these parameters, it's impossible to imagine galaxies forming before a billion years. The Hubble Space Telescope was already able to obtain images in the near infrared. Early images of distant objects appeared to show groups of mini-galaxies. But the James Webb Space Telescope showed that these objects were nothing other than HII regions belonging to barred spiral galaxies, fully formed, hosting old stars, only 500 million years old.\\

For decades, the Standard Model $\Lambda$CDM has relied on its ability to account for CMB fluctuations as gravito-acoustic oscillations, by adapting the numerous parameters relating to dark matter, dark energy and, in particular, the value of the Hubble constant. This desire to match observational data has resulted in a Hubble constant value of \qty{67}{\kilo\meter\per\second\per\mega\parsec}. This is significantly lower than the value of \qty{70}{\kilo\meter\per\second\per\mega\parsec} deduced from direct observation of standard candles.\\

All these factors are creating a deep crisis within the specialist community, and some voices are beginning to be heard, suggesting the need to consider a profound paradigm shift. This is what the Janus cosmological model\footnote{see \vref{JMC}, where this model is developped.} proposes.\\

Since we are unable to introduce negative masses into the general relativity model, let's consider a profound change of geometric paradigm, already evoked in the previous sections under the aspect of group theory and topology. The motion of positive masses, immersed in the gravitational field, takes place according to geodesics that we consider to be derived from a first metric $\gmunu$. We will therefore describe the motion of negative masses using a second
set of geodesics, derived from a second metric $\gbarmunu$. We thus have a manifold, whose points are marked by the coordinates $(x^0,x^1,x^2,x^3)$, equipped with a pair of metrics $(\gmunu,\gbarmunu)$. We shall neglect the action of electromagnetic fields and consider only the field of gravity. From the metrics and we can construct Ricci tensors $\Rmunu$ and $\Rbarmunu$ and their associated Ricci scalars $R$ and $\Rbar$.\\

As the Janus model proposes a paradigm shift by introducing a bimetric structure to account for both positive and negative masses, we will now focus on the foundational mathematical structure underlying this model. The Janus cosmological model builds upon the interaction between two entities, i.e. positive and negative mass populations, each associated with its respective metric. In the following section, we will explore the formulation of the action and field equations governing this interaction, and how these coupled systems lead to a coherent description of cosmic phenomena, offering an alternative to the limitations of the standard model $\Lambda$CDM.

\section{Foundation of the Janus cosmological model}\label{JMC}

To build this model, let us now consider the interaction between two entities: ordinary matter with positive mass interacting with negative mass through gravitational effects. This model involving negative mass takes into account the influence of both dark matter and dark energy.\\

We can describe this system of two entities with respective metrics \( g_{\mu\nu} \) and \( \gbar_{\mu\nu} \). Let \( R \) and \( \Rbar \) be the corresponding Ricci scalars. We then consider the following two-layer action\footnote{Integration over $\mathscr{E}$ using the element $\mathrm{d}^4x$ is a method for computing the total action in the bimetric spacetime, reflecting the four-dimensional nature of this bimetric universe. This implies considering the entire spacetime as the domain of integration, integrating the contributions from each point to the action. The term $\mathrm{d}^4x$ represents an infinitesimal element of hypervolume of this bimetric spacetime, used to measure each segment during integration. Thus, it is a multiple volume integral performed over the four dimensions of spacetime, accumulating contributions to the total action from each four-dimensional volume segment corresponding to each metric.}:
\begin{equation}
A = \int_{\mathscr{E}} \left( \frac{1}{2\chi} R + S + \mathcal{S} \right) \sqrt{|g|} \, \mathrm{d}^4x + \int_{\mathscr{E}} \left( \frac{\kappa}{2\bar{\chi}} \Rbar + \Sbar + \mathcal{\Sbar} \right) \sqrt{|\gbar|} \, \mathrm{d}^4x.
\label{chapter3_eq9}
\end{equation}

The terms \( S \) and \( \Sbar \) will give the source terms related to the populations of the two entities, while the terms \( \mathcal{S} \) and \( \mathcal{\Sbar} \) will generate the interaction tensors. \( \chi \) and \( \bar{\chi} \) are the Einstein gravitational constants for each entity. \( g \) and \( \gbar \) are the determinants of the metrics \( g_{\mu\nu} \) and \( \gbar_{\mu\nu} \). For \( \kappa = \pm 1 \), we apply the principle of least action. The Lagrangian derivation of this action gives us:

\begin{equation}
\begin{aligned}
0 &= \delta A ,\\
&= \int_{\mathscr{E}} \delta \left( \frac{1}{2\chi} R + S + \mathcal{S} \right) \sqrt{|g|} \, \mathrm{d}^4 x \\
&\quad + \int_{\mathscr{E}} \delta \left( \frac{\kappa}{2\bar{\chi}} \Rbar + \Sbar + \mathcal{\Sbar} \right) \sqrt{|\gbar|} \, \mathrm{d}^4 x ,\\
&= \int_{\mathscr{E}} \delta \left[ \frac{1}{2\chi} \left( \frac{\delta R}{\delta {{g^{\mu\nu}}}} + \frac{R}{\sqrt{|g|}} \frac{\delta\sqrt{|g|}}{\delta {g^{\mu\nu}}} \right) \right.\\
&\quad\left. + \frac{1}{\sqrt{|g|}} \frac{\delta(\sqrt{|g|} S)}{\delta {g^{\mu\nu}}} + \frac{1}{\sqrt{|g|}} \frac{\delta(\sqrt{|g|} \mathcal{S})}{\delta {g^{\mu\nu}}} \right] \delta {g^{\mu\nu}} \sqrt{|g|} \, \mathrm{d}^4 x \\
&\quad + \int_{\mathscr{E}} \delta \left[ \frac{\kappa}{2\bar{\chi}} \left( \frac{\delta \Rbar}{\delta {\gbar^{\mu\nu}}} + \frac{\Rbar}{\sqrt{|\gbar|}} \frac{\delta\sqrt{|\gbar|}}{\delta {\gbar^{\mu\nu}}} \right) \right.\\
&\quad\left. + \frac{1}{\sqrt{|\gbar|}} \frac{\delta(\sqrt{|\gbar|} \Sbar)}{\delta {\gbar^{\mu\nu}}} + \frac{1}{\sqrt{|\gbar|}} \frac{\delta(\sqrt{|\gbar|} \mathcal{\Sbar})}{\delta {\gbar^{\mu\nu}}} \right] \delta {\gbar^{\mu\nu}} \sqrt{|\gbar|} \, \mathrm{d}^4 x .
\end{aligned}
\label{chapter3_eq10}
\end{equation}

For any variation \( \delta {g^{\mu\nu}} \) and \( \delta {\gbar^{\mu\nu}} \), we locally obtain:

\begin{equation}
\frac{1}{2\chi} \left( \frac{\delta R}{\delta {g^{\mu\nu}}} + \frac{R}{\sqrt{|g|}} \frac{\delta\sqrt{|g|}}{\delta {g^{\mu\nu}}} \right) + \frac{1}{\sqrt{|g|}} \frac{\delta(\sqrt{|g|} S)}{\delta {g^{\mu\nu}}} + \frac{1}{\sqrt{|g|}} \frac{\delta(\sqrt{|g|} \mathcal{S})}{\delta {g^{\mu\nu}}} = 0 ,
\label{chapter3_eq11a}
\end{equation}
\begin{equation}
\frac{\kappa}{2\bar{\chi}} \left( \frac{\delta \Rbar}{\delta {\gbar^{\mu\nu}}} + \frac{\Rbar}{\sqrt{|\gbar|}} \frac{\delta\sqrt{|\gbar|}}{\delta {\gbar^{\mu\nu}}} \right) + \frac{1}{\sqrt{|\gbar|}} \frac{\delta(\sqrt{|\gbar|} \Sbar)}{\delta {\gbar^{\mu\nu}}} + \frac{1}{\sqrt{|\gbar|}} \frac{\delta(\sqrt{|\gbar|} \mathcal{\Sbar})}{\delta {\gbar^{\mu\nu}}} = 0 .
\label{chapter3_eq11b}
\end{equation}
Let us then introduce the following tensors:
\begin{align}
T_{\mu\nu} &= -\frac{2}{\sqrt{|g|}} \frac{\delta(\sqrt{|g|} S)}{\delta {g^{\mu\nu}}} = -2 \frac{\delta S}{\delta {g^{\mu\nu}}} + g_{\mu\nu} S ,\label{chapter3_eq12}\\
\overline{T}_{\mu\nu} &= -\frac{2}{\sqrt{|\gbar|}} \frac{\delta(\sqrt{|\gbar|} \Sbar)}{\delta {\gbar^{\mu\nu}}} = -2 \frac{\delta \Sbar}{\delta {\gbar^{\mu\nu}}} + \gbar_{\mu\nu} \Sbar , \label{chapter3_eq13}\\
\mathcal{T}_{\mu\nu} &= -\frac{2}{\sqrt{|\gbar|}} \frac{\delta(\sqrt{|g|} \mathcal{S})}{\delta {g^{\mu\nu}}} , \label{chapter3_eq14}\\
\overline{\mathcal{T}}_{\mu\nu} &= -\frac{2}{\sqrt{|g|}} \frac{\delta(\sqrt{|\gbar|} \mathcal{\Sbar})}{\delta {\gbar^{\mu\nu}}} . \label{chapter3_eq15}
\end{align}
We obtain then from equations (\ref{chapter3_eq14}) and (\ref{chapter3_eq15}):
\begin{equation}
\sqrt{\frac{|\gbar|}{|g|}} \mathcal{T}_{\mu\nu} = \sqrt{\frac{|\gbar|}{|g|}} \frac{-2}{\sqrt{|\gbar|}} \frac{\delta(\sqrt{|g|} \mathcal{S})}{\delta {g^{\mu\nu}}} = \frac{-2}{\sqrt{|g|}} \frac{\delta(\sqrt{|g|} \mathcal{S})}{\delta {g^{\mu\nu}}} = - 2 \frac{\delta \mathcal{S}}{\delta {g^{\mu\nu}}} + g_{\mu\nu} \mathcal{S} ,
\label{chapter3_eq32}
\end{equation}
\begin{equation}
\sqrt{\frac{|g|}{|\gbar|}} \overline{\mathcal{T}}_{\mu\nu} = \sqrt{\frac{|g|}{|\gbar|}} \frac{-2}{\sqrt{|g|}} \frac{\delta(\sqrt{|\gbar|} \mathcal{\Sbar})}{\delta {\gbar^{\mu\nu}}} = \frac{-2}{\sqrt{|\gbar|}} \frac{\delta(\sqrt{|\gbar|} \mathcal{\Sbar})}{\delta {\gbar^{\mu\nu}}} = - 2 \frac{\delta \mathcal{\Sbar}}{\delta {\gbar^{\mu\nu}}} + \gbar_{\mu\nu} \mathcal{\Sbar} .
\label{chapter3_eq33}
\end{equation}
Introduced into equations (\ref{chapter3_eq11a}) and (\ref{chapter3_eq11b}), we can thus deduce the coupled field equations describing the system of the two entities:
To obtain the desired interaction laws under the Newtonian approximation, we must choose \( \kappa = -1 \). The system of equations then becomes:
\begin{equation}
R_{\mu\nu} - \frac{1}{2} g_{\mu\nu} R = \chi \left(T_{\mu\nu} + \sqrt{\frac{|\gbar|}{|g|}} \mathcal{T}_{\mu\nu}\right) ,
\label{chapter3_eq34_general}
\end{equation}
\begin{equation}
\Rbar_{\mu\nu} - \frac{1}{2} \gbar_{\mu\nu} \Rbar = \kappa\bar{\chi} \left(\overline{T}_{\mu\nu} + \sqrt{\frac{|g|}{|\gbar|}} \overline{\mathcal{T}}_{\mu\nu} \right) .
\label{chapter3_eq35_general}
\end{equation}

To obtain the desired interaction laws under the Newtonian approximation, we must choose \( \kappa = -1 \). The system of equations then becomes:
\begin{equation}
R_{\mu\nu} - \frac{1}{2} g_{\mu\nu} R = \chi \left(T_{\mu\nu} + \sqrt{\frac{|\gbar|}{|g|}} \mathcal{T}_{\mu\nu}\right) ,
\label{chapter3_eq34}
\end{equation}
\begin{equation}
\Rbar_{\mu\nu} - \frac{1}{2} \gbar_{\mu\nu} \Rbar = -\bar{\chi} \left(\overline{T}_{\mu\nu} + \sqrt{\frac{|g|}{|\gbar|}} \overline{\mathcal{T}}_{\mu\nu} \right) .
\label{chapter3_eq35}
\end{equation}
The tensor \( T_{\mu\nu} \) is the energy-momentum tensor, which represents the source of the field acting on positive mass entities and positive-energy photons. The term $\sqrt{\frac{|\gbar|}{|g|}}$ is the source of this field attributed to the action of negative masses on these positive masses. The tensor \( \overline{T}_{\mu\nu} \) is the energy-momentum tensor, which represents the source of the field acting on negative mass entities and negative-energy photons, and the term $\sqrt{\frac{|g|}{|\gbar|}}$ is the source of this field attributed to the action of positive masses on these negative masses. \( \mathcal{T}_{\mu\nu} \) and \( \overline{\mathcal{T}}_{\mu\nu} \) are the interaction tensors of the system of the two entities corresponding to the \textit{"induced geometry"}, meaning how each matter distribution on one layer of the universe contributes to the geometry of the other\footnote{Interaction between populations of positive and negative masses.}.\\

General relativity produces only a limited number of exact solutions. We will follow the same logic.\\

Having established the foundation of the Janus cosmological model, with its bimetric structure and the corresponding field equations, we turn to constructing explicit solutions under the assumption of homogeneity and isotropy. By considering the FLRW form for both metrics, we aim to derive a time-dependent solution that accounts for the interaction between positive and negative mass populations. This section will focus on obtaining these solutions, exploring their compatibility with observational data, and providing a theoretical framework for the accelerated cosmic expansion.

\section{Construction of a time-dependent, homogeneous and isotropic solution}

Given the symmetry assumptions, the metrics then have the FLRW form. The variable $x^0$ is the
common chronological coordinate (time marker).

\begin{subequations}
	\begin{align}
		\gmunu&=\di {x^0}^2-a^2\left[\frac{\di u^2}{1-ku^2}+u^2\di\theta^2+u^2\sin^2\theta\di\varphi^2\right],\\
		\gbarmunu&=\di {x^0}^2-\abar^2\left[\frac{\di u^2}{1-\kbar u^2}+u^2\di\theta^2+u^2\sin^2\theta\di\varphi^2\right].
	\end{align}
\end{subequations}
The determinants of the two metrics are
\begin{equation}
	g=-a^6\sin^2\theta, \qquad \gbar=-\abar^6\sin^2\theta.
\end{equation}
As shown in reference \cite{Petit2014} the treatment of the two equations leads to the compatibility relation:
\begin{equation}
	\rho c^2a^3+\rhobar \cbar^2\abar^3=E=\text{cst}.
\label{eqConservEnergie}
\end{equation}
This translates into conservation of energy, extended to both populations. The exact solution, referring to two dust universes, corresponds to:
\begin{equation}
	k=\kbar=-1
\end{equation}
and:

\begin{subequations}
	\begin{align}
		a^2\frac{\di^2 a}{\di {x^0}^2}&=-\frac{4\pi G}{c^2}E,\\
		\abar^2\frac{\di^2 \abar}{\di {x^0}^2}&=+\frac{4\pi G}{\cbar^2}E.
	\end{align}
\end{subequations}

\begin{figure}
	\centering
	\includegraphics[width=11cm]{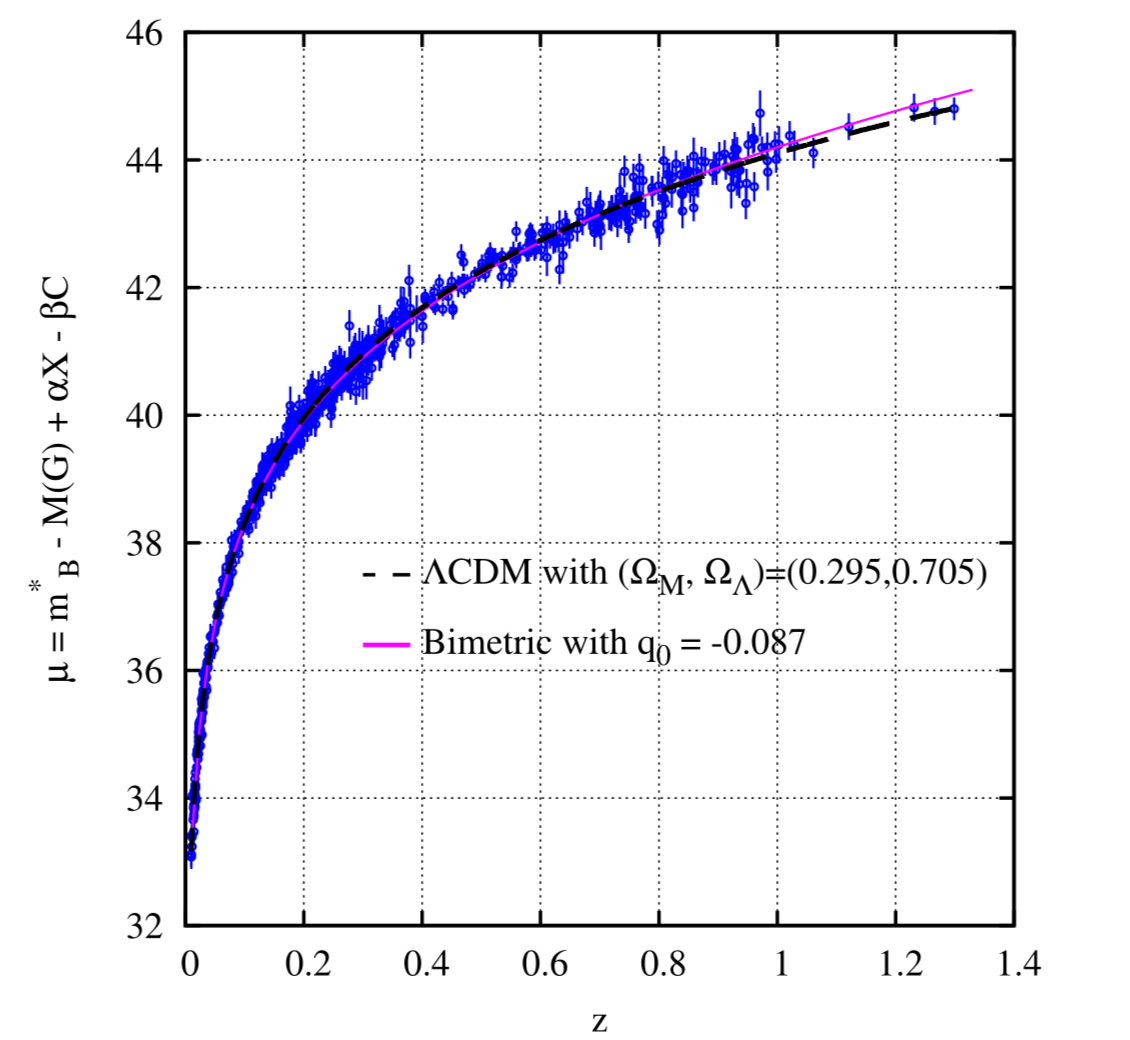}
	\caption{Comparison of observed and theoretical magnitudes as a function of $z$ redshift~\cite{DAgostini2018}.}
	\label{magnitude}
\end{figure}

A theoretical model loses interest if it cannot be compared with observational data. The evolution of the positive species will correspond to an acceleration if the energy $E$ of the system is negative. This provides a physical interpretation of the acceleration of the cosmic expansion (\cite{Perlmutter1999, Riess1998}), which then follows from the fact that the energy content is predominantly negative. Numerical data have been successfully compared with observational data \cite{DAgostini2018}. The corresponding curve is shown in \vref{magnitude}.\\

To complete the model, we now need to provide exact stationary solutions. We will restrict ourselves to $\mathfrak{so}(3)$ symmetry.\\

We now focus our attention to the interaction laws and their observational consequences. These interaction laws, derived from the coupled field equations of the Janus model, govern how positive and negative mass entities influence each other. The next section explores these laws in detail and examines how they provide explanations for various cosmological phenomena, including the formation of large-scale structures and the resolution of issues related to dark matter and dark energy.

\section{Interaction laws and observational consequences}

In the system of coupled field equations (\ref{chapter3_eq34}) and (\ref{chapter3_eq35}), the terms on the left-hand side involve the Ricci tensors \( R_{\mu\nu} \) and \( \Rbar_{\mu\nu} \) and the corresponding Ricci scalars \( R \) and \( \Rbar \). These terms are calculated from the two metrics \( g_{\mu\nu} \) and \( \gbar_{\mu\nu} \). 
Using these two metrics, we then calculate the form of two operators known as \emph{covariant derivatives} \( \nabla_{\mu} \) and \( \upnablabar_{\mu} \). It turns out that, due to their form, the two left-hand sides of both equations identically satisfy the following relation:
\begin{align}
\nabla_{\mu} \left( R_{\mu\nu} - \frac{1}{2} R g_{\mu\nu} \right) &= 0, \\
\upnablabar_{\mu} \left( \Rbar_{\mu\nu} - \frac{1}{2} \Rbar \gbar_{\mu\nu} \right) &= 0.
\end{align}
%The two tensors, \( T_{\mu\nu} \) and \( \overline{T}_{\mu\nu} \), also satisfy the following condition:
The corresponding covariant derivatives of the two second members must therefore also be zero, which corresponds to the Bianchi identities, implying:
\begin{align}
\nabla_{\mu} T_{\mu\nu} = 0 \label{eqCov-T}, \\
\upnablabar_{\mu} \Tbarmunu = 0 \label{eqCov-Tbar}.
\end{align}
We should also have:
%\begin{align}
%\nabla_{\mu} \left[\sqrt{\lfrac[2mu]{\gbar}{g}}\mathcal{T}_{\mu\nu}\right]=0 \label{eqCov-sqrt-t}, \\
%\upnablabar_{\mu} \left[\sqrt{\lfrac[2mu]{\mathstrut g}{\gbar}}\overline{\mathcal{T}}_{\mu\nu}\right]=0. \label{eqCov-sqrt-tbar}
%\end{align}
\begin{align}
\nabla_{\mu} \left[\sqrt{\displaystyle \frac{\gbar}{g}}\mathcal{T}_{\mu\nu}\right]=0 \label{eqCov-sqrt-t}, \\
\upnablabar_{\mu} \left[\sqrt{\displaystyle \frac{g}{\gbar}}\overline{\mathcal{T}}_{\mu\nu}\right]=0. \label{eqCov-sqrt-tbar}
\end{align}

In stationary conditions, the square roots of the ratios of the determinants behave like constants, reflecting an \textit{``apparent mass effect''}. Conditions \eqref{eqCov-sqrt-t} and \eqref{eqCov-sqrt-tbar} can therefore be replaced by:

\begin{equation}
	\nabla_{\mu} \mathcal{T}_{\mu\nu}=0,
\label{eqCov-t}	
\end{equation}
and

\begin{equation}
	\upnablabar_{\mu} \overline{\mathcal{T}}_{\mu\nu}=0.
\label{eqCov-tbar}	
\end{equation}

Let's write the system of equations in mixed notation, replacing the square roots, which have become constant, by the positive constants $b^2$ and $\bbar^2$:
\begin{subequations}
	\begin{align}
		\Rmixmunu-\frac{1}{2}R\gmixmunu&=\chi\left[\Tmixmunu+b^2\tmixmunu\right],\\
		\Rbarmixmunu-\frac{1}{2}\Rbar\gbarmixmunu&=-\chi\left[\Tbarmixmunu+\bbar^2\tbarmixmunu\right].
	\end{align}
	\end{subequations}
Using the Newtonian approximation, in both populations the non-zero tensor terms reduce to:
\begin{equation}
	\Tmixzz=\rho c^2>0 \qquad \tmixzz=\rhobar\cbar^2<0 \qquad \Tbarmixzz=\rhobar\cbar^2<0 \qquad \tbarmixzz=\rho c^2>0.
\end{equation}
In our system of coupled field equations, the presence of a minus sign in front of the second member of the second equation gives the following interaction laws:

\begin{itemize}
	\item Masses of the same sign attract each other;
	\item Masses of opposite signs repel each other.
\end{itemize}

We have thus eliminated the runaway effect.\\

The first conclusion to be drawn is that where one of the two types of mass is present, the other is absent, as immediately confirmed by simulations \cite{Petit1995}. This is the case in the vicinity of the Sun, and under these conditions the first equation is identified with Einstein's 1915 equation. The model is therefore in line with all the classical local observational data of general relativity: Mercury's perihelion advance, deflection of light rays by the Sun. The model therefore does not invalidate that of general relativity, but presents itself as its extension, made essential to integrate the new observational data, which can no longer be managed by introducing the hypothetical components of dark matter and dark energy.\\

\begin{figure}[h!]
\centering
\includegraphics[width=\textwidth]{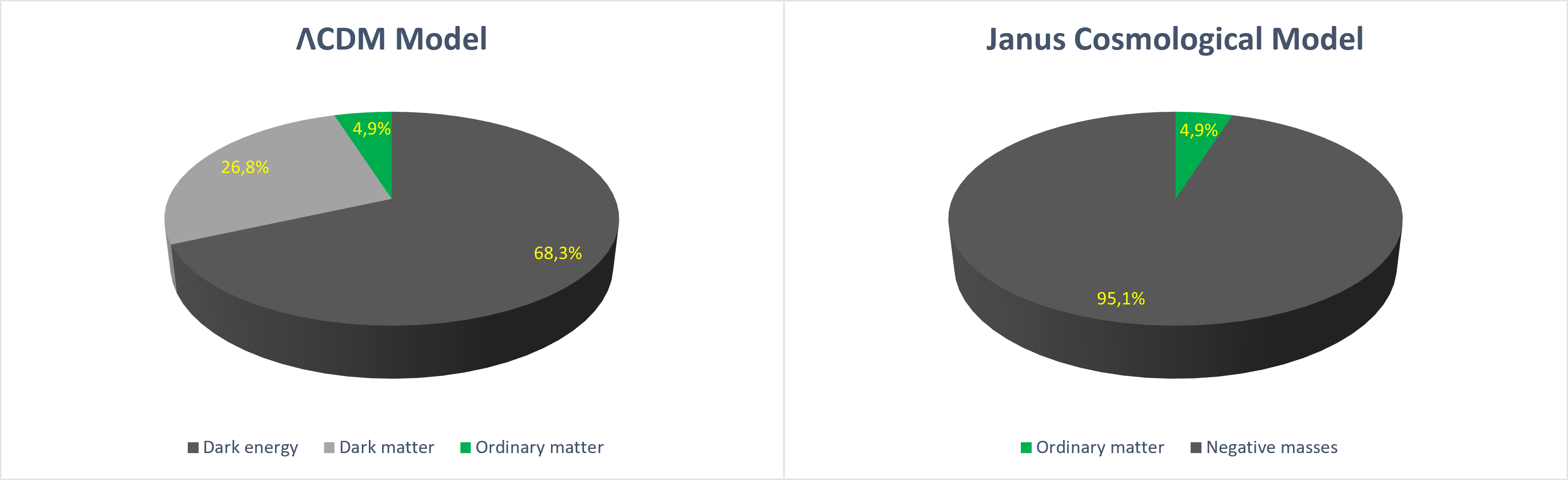} % Ensure correct path and extension
\caption{Comparative contents of the $\Lambda$CDM and Janus models.}
\label{LCDM-JCM}
\end{figure}

We have seen, in our construction of the unsteady solution, that negative energy dominates. The model is thus profoundly asymmetrical. The negative mass component is proposed as a substitute for the combined roles traditionally attributed to dark matter and dark energy. By the way, going back to the original idea, inspired by the work of Andreï Sakharov, this allows us to attribute a well-defined identity to these components. They are invisible, insofar as negative masses emit photons of negative energy that our optical instruments cannot capture. They are therefore simply copies of our own antimatter, assigned a negative mass. We then have a new distribution of contents (see \vref{LCDM-JCM}).\\

At the moment of decoupling, when the gravitational instability can play its role (we must then speak of joint gravitational instabilities), the characteristic Jeans time is shorter for negative masses:

\begin{equation}
	\tJbar=\frac{1}{\sqrt{4\pi G\abs{\rhobar}}}\ll t_\text{J}=\frac{1}{\sqrt{4\pi G\rho}}.
\end{equation}

The result will be a regular distribution of negative-mass conglomerates of spheroidal antihydrogen and negative-mass antihelium. These will behave like immense negative-mass protostars. As soon as their temperature causes hydrogen reionization, their contraction will cease. These formations will then radiate in the red and infrared wavelengths. But their cooling time is then large compared to the age of the universe, which means that these objects will no longer evolve. The history of this universe fold associated with negative masses is totally different from our universe fold of ordinary matter. It will not give rise to stars, galaxies or planets. It will contain no atoms heavier than negative-mass antihelium. And there will be no life. And, as we'll see later: these negative formations are deliberately situated within the Newtonian approximation.\\

\begin{figure}
	\centering
	\includegraphics[width=13cm]{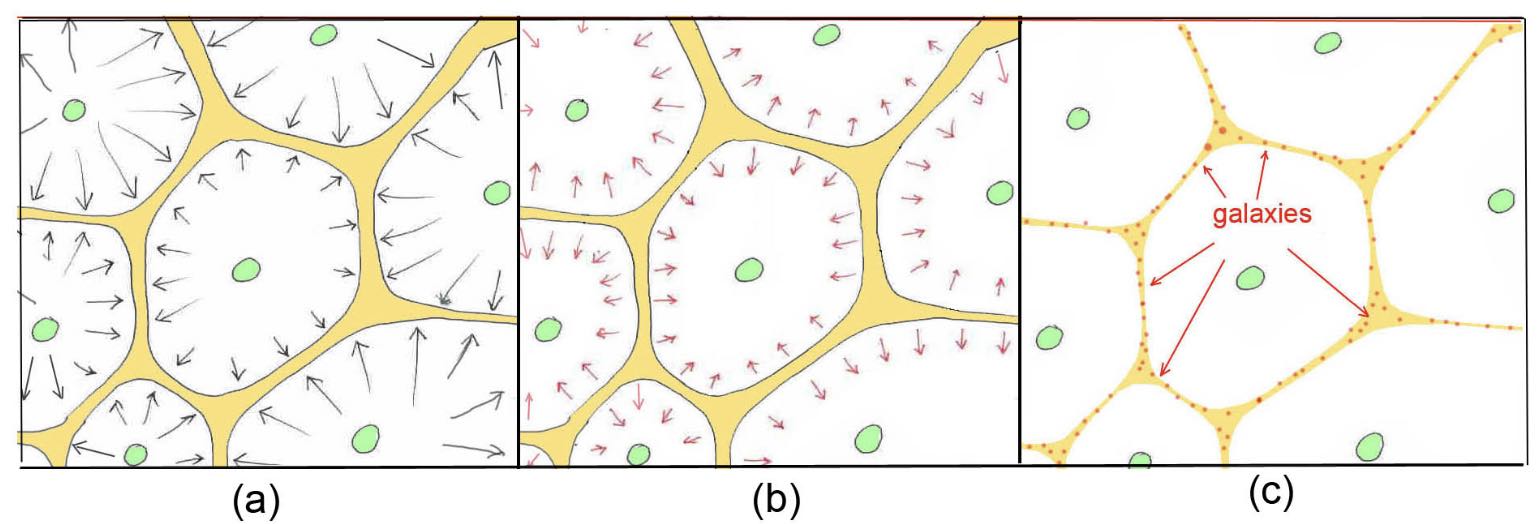}
	\caption{Early rapid star and galaxy formation.}
	\label{galaxy-formation}
\end{figure}

But there's another very important point. When these spheroidal conglomerates form, they confine the positive mass to the residual space, giving it a lacunar structure, comparable to joined soap bubbles. The negative mass is thus distributed in the form of thin plates,
sandwiched between two negative conglomerates that exert a strong back pressure on it. The positive mass is thus violently compressed and heated. However, due to its plate-like arrangement, it can cool down very quickly through the emission of radiation (see \vref{galaxy-formation}).\\

The result is a pattern of first-generation star and galaxy formation totally different from the standard one. This configuration had been the subject of simulations \cite{Petit1995} since the first, heuristic, approach to the model, and the fact that objects all form within the first hundred million years was one of its predictions, largely confirmed by JWST data.\\

\begin{figure}
	\centering
	\includegraphics[width=11cm]{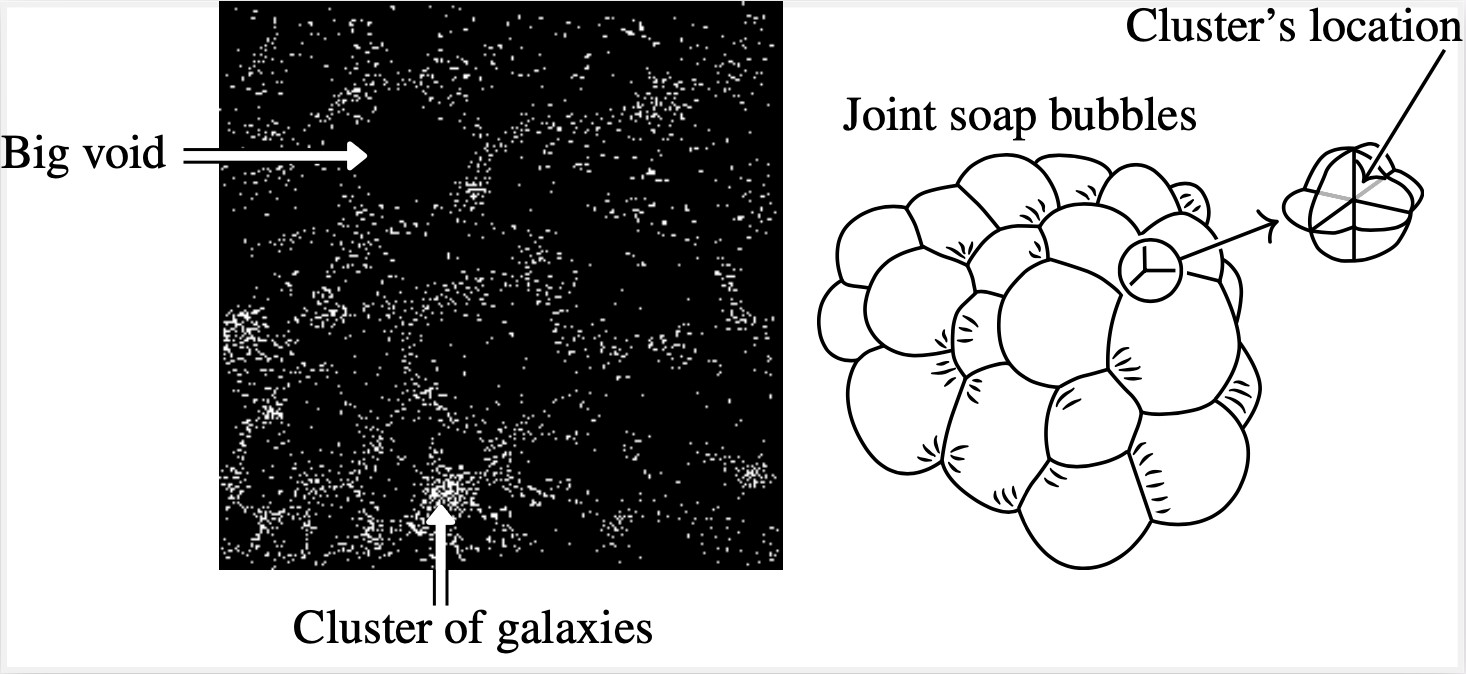}
	\caption{Structure of positive mass in contiguous bubbles.}
	\label{bubbles}
\end{figure}

The lacunar structure, advocated as early as 1995 \cite{Petit1995}, predicted the existence of large voids, which the discoveries of the dipole repeller and other similar large voids have also confirmed. Once this lacunar structure has been formed, matter tends to concentrate along the segments common to three gaps, forming filaments (see \vref{bubbles}). The nodes of this distribution will only develop into galaxy clusters.\\

After establishing the interaction laws and exploring their observational consequences, it's essential to verify the mathematical and physical consistency of the Janus model. This requires demonstrating that the system of coupled field equations respects the Bianchi identities and provides consistent solutions in the weak field limit. In the following section, we will examine the conditions necessary to ensure this consistency, particularly in regions dominated by ordinary matter, such as near the Sun, as well as in regions dominated by negative masses, such as near the dipole repeller.

\section{The mathematical and physical consistency of the model}

This is ensured in an isotropic, homogeneous and unsteady situation, the required condition being the generalized conservation of energy expressed by equation \eqref{eqConservEnergie}. We now turn to the case of stationary solutions, limiting ourselves to those that satisfy \(\mathfrak{so}(3)\) symmetry. Bianchi identities must then be satisfied, i.e.~relations \eqref{eqCov-T}, \eqref{eqCov-Tbar}, \eqref{eqCov-t} and \eqref{eqCov-tbar}.\\

First, we'll show the existence of asymptotic consistency in Newtonian approximation situations. The key aspects of this approximation are as follows:

\begin{itemize}
	\item Velocities must be negligible compared to the speed of light. This is the case for velocities $\langle v \rangle$ and $\langle \vbar \rangle$ of thermal agitation in both media, which are involved in the definition of
pressures and in both media. After decoupling:

\begin{equation}
	\varepsilon p=\frac{\varepsilon\rho\langle v \rangle}{3} \quad\text{and}\quad  \varepsilon \pbar=\frac{\varepsilon\rhobar\langle \vbar \rangle}{3}.
\end{equation}

\item Curvature effects must be neglected, meaning that the radial coordinate must be much larger than the characteristic length scale associated with curvature, i.e., the Schwarzschild radius.
\end{itemize}

\subsection{Newtonian approximation of the field generated by a positive mass M}

Let's introduce the Schwarzschild radius \(\Rs\) as follows:
\begin{equation}
	\varepsilon\Rs=\varepsilon\frac{2GM}{c^2},
\end{equation}
where $\varepsilon$ being a small parameter. $\mathfrak{so}(3)$ symmetry imposes the shapes of the two metrics:

\begin{subequations}
	\begin{align}
		\di s^2&=\mathrm{e}^\nu\di {x^0}^2-\mathrm{e}^{\lambda}\di r^2-r^2\di\theta^2-r^2\sin^2\theta\di\varphi^2,\label{eq-ds2}\\
		\di \sbar^2&=\mathrm{e}^{\nubar} \di {x^0}^2-\mathrm{e}^{\lambdabar}\di r^2-r^2\di\theta^2-r^2\sin^2\theta\di\varphi^2.\label{eq-dsbar2}
	\end{align}
\end{subequations}

The construction of a stationary solution then requires to calculate the functions:

\begin{equation}
	\nu(r),\quad \lambda(r),\quad \nubar(r),\quad\text{and}\quad \lambdabar(r).
\end{equation}
To locate this solution, we need to consider the shapes of the field source tensors:

\begin{equation}
	\Tmixmunu, \quad \tmixmunu, \quad \Tbarmixmunu, \quad\text{and}\quad \tbarmixmunu.
\end{equation}

Let's start by considering a situation where only positive mass is present. The tensors $\tmixmunu$ and
$\Tbarmixmunu$ are then null and the two field equations \ref{chapter3_eq34} and \ref{chapter3_eq35} become in mixed-mode form:
\begin{align}
\Rmixmunu - \frac{1}{2} \gmixmunu R &= \chi \Tmixmunu , \label{eq_positivemass1} \\
\Rbarmixmunu - \frac{1}{2} \gbarmixmunu \Rbar &= -\bar{\chi} \sqrt{\frac{|g|}{|\gbar|}} \tbarmixmunu. \label{eq_positivemass2}
\end{align}
The form of the tensor $\Tmixmunu$ in its classical mixed-mode form is given by\footnote{(13.1) p.425 of \cite{Adler1975}.}:

\begin{equation}
	\Tmixmunu=\begin{pmatrix}
		\rho c^2 & 0 & 0 & 0 \\
		0 & -\varepsilon p & 0 & 0 \\
		0 & 0 & -\varepsilon p & 0 \\
		0 & 0 & 0 & -\varepsilon p
	\end{pmatrix}.
\label{eq-Tmunu}
\end{equation}
As we are in the Newtonian approximation, $\varepsilon$ is very small. With the introduction of the metric \eqref{eq-ds2} and the tensor \eqref{eq-Tmunu} in the first field equation, we are led to introduce the function $m(r)$ such that:

\begin{equation}
\mathrm{e}^{-\lambda} = 1 - \frac{2m(r)}{r} \implies 2m(r) = r \left( 1 - \mathrm{e}^{-\lambda} \right).
\end{equation}

%with :
%\begin{equation}
%m = \frac{GM}{c^2} 
%\label{chapter3_eq_variable_change_pos_m}
%\end{equation}

Similarly to equation (14.18) from \cite{Adler1975}, the classic calculation leads to the relationship:
\begin{equation}
m(r) = \frac{G\rho}{c^2} \int_{0}^{r} 4\pi r^2 \mathrm{d}r = \frac{4}{3} \pi r^3 \rho \frac{G}{c^2}.
\label{eq-m(r)}
\end{equation}
%\begin{equation}
%	m(r)=2\frac{4\pi Gr^2\rho}{3c^2}\leq\frac{2GM}{c^2}=\Rs.
%\end{equation}
We then obtain the classical Tolman--Oppenheimer--Volkoff (TOV) equation (\cite{Oppenheimer1939}). Relation \eqref{eq-m(r)} places the small quantity in front of any quantity that will be neglected in the Newtonian approximation:
\begin{equation}
\frac{1}{{c}^2} \frac{\di p}{\di r} = -\frac{{m} + \frac{4\pi\varepsilon G {p} r^3}{{c}^4}}{r (r - 2{m\varepsilon})} \left( {\rho} + \varepsilon\frac{{p}}{{c}^2} \right).
\label{eq-dp/dr}
\end{equation}
When $\varepsilon$ tends to zero (or $c$ tends to infinity) we get:
\begin{equation}
	\frac{\di p}{\di r}=-\frac{\rho mc^2}{r^2}=-\frac{G\rho}{r^2}\frac{4\pi r^3\rho}{3}.
\end{equation}
The quantity $\frac{4\pi r^3\rho}{3}$ represents the amount of matter $\mu(r)$ contained inside a sphere of
radius~$r$. We know that the force of gravity exerted inside a mass of constant density is
 equivalent to that exerted by the mass located at the center of the sphere, and that the mass
located outside this sphere gives a force of zero. So the quantity $-\frac{G\rho\mu(r)}{r^2}$ is the force of
gravity, per unit volume, acting on the matter contained in an elementary volume around a point at distance $r$ from the center. Thus the relation \eqref{eq-Tmunu}, which follows from the Newtonian approximation, expresses that the force of gravity balances the force of pressure. This is the classic Euler relationship.\\

Hence, the Schwarzschild interior metric built is given by:
\begin{equation}
{\mathrm{d}s}^{2} = \left[ \frac{3}{2} \sqrt{\left( 1 - \frac{{r_n}^{2}}{{\hat{r}}^{2}} \right)} - \frac{1}{2} \sqrt{\left( 1 - \frac{r^{2}}{{\hat{r}}^{2}} \right)} \right]^{2} \mathrm{d}x^{0^{2}} - \frac{\mathrm{d}r^{2}}{1 - \frac{r^{2}}{{\hat{r}}^{2}}} - r^{2} \left( \mathrm{d}\theta^{2} + \sin^{2}\theta \mathrm{d}\phi^{2} \right).
\label{chapter3_eq_tov7_1}
\end{equation}
This metric connects with the Schwarzschild exterior metric :
		\begin{equation}
		{\mathrm{d}s}^{2} = \left(1 - \frac{2 G M}{c^2 r}\right) c^2 \mathrm{d}x^{0^{2}}
		- \frac{\mathrm{d}r^2}{1 - \frac{2 G M}{c^2 r}} 
		- r^2 \left( \mathrm{d}\theta^2 + \sin^2\theta \mathrm{d}\phi^2 \right),
		\label{chapter3_eq_tov7_1_ext}
		\end{equation}
where $r_n$ is the radius of the star and $\hat{r}$ is a stellar constant as a function of its density $\rho$. It's the characteristic radius of a neutron star, defined under the assumption of constant density $\rho$. It establishes a critical threshold for the star's radius, beyond which the internal pressure becomes infinite at the center, indicating a physical singularity or instability. This radius is derived from the balance between gravitational forces and the internal pressure gradients within the star (\cite{aop2024}). It's given by:
\begin{equation}
\hat{r} = \sqrt{\frac{3c^2}{8\pi G \rho}}.
\label{chapter5_eq2}
\end{equation}
We can thus deduce, according to the classical theory of general relativity, that a particle of ordinary matter will undergo an attractive gravitational field due to the effect of a distribution of positive masses.\\

To ensure the mathematical consistency of the system of two field equations \ref{eq_positivemass1} and \ref{eq_positivemass2}, we therefore need to consider a form of the tensor $\tbarmixmunu$ that gives back this same Euler relation when the
Newtonian approximation is also applied to this solution. This is guaranteed with the form of the interaction tensor $\tbarmixmunu$ of the field equation \ref{eq_positivemass2} as this choice can stem from a Lagrangian derivation:
\begin{equation}
	\tbarmixmunu=\begin{pmatrix}
		\rho c^2 & 0 & 0 & 0 \\
		0 & +\varepsilon p & 0 & 0 \\
		0 & 0 & +\varepsilon p & 0 \\
		0 & 0 & 0 & +\varepsilon p
	\end{pmatrix}.
\end{equation}

On the right-hand side of the second field equation (\ref{eq_positivemass2}), the ratio of determinants will be considered almost unity insofar as we perform this calculation within the Newtonian approximation\footnote{In stationary conditions, the square roots of the ratios of the determinants behave like constants, reflecting an \textit{``apparent mass effect''}.}.\\

Then, if we consider that:
\begin{equation}
\sqrt{\frac{|g|}{|\gbar|}} = \sqrt{\frac{\mathrm{e}^\nu \mathrm{e}^\lambda r^4 \sin^2 \theta}{{\mathrm{e}^{\bar{\nu}} \mathrm{e}^{\bar{\lambda}} r^4 \sin^2 \theta}}} \approx 1,
\end{equation}
the calculation leads to the Tolman--Oppenheimer--Volkoff (TOV) solution for the population of negative masses managed by the second field equation:
\begin{equation}
	\frac{1}{{c}^2} \frac{\di p}{\di r} = -\frac{{m} - \frac{4\pi\varepsilon G {p} r^3}{{c}^4}}{r (r + 2{m\varepsilon})} \left( {\rho} - \varepsilon\frac{{p}}{{c}^2} \right).
\label{eq-dp/dr2}	
\end{equation}
The two solutions, \eqref{eq-dp/dr} and \eqref{eq-dp/dr2}, asymptotically approach the Euler equation in the Newtonian approximation as \( \varepsilon \) tends to zero. This also corresponds to the asymptotic satisfaction of the Bianchi identities in the same context\footnote{The inequality \(r \gg 2m\) (where \(m\) is often replaced by \( \frac{GM}{c^2} \) to obtain a dimension of length, \(M\) being the mass of the object and \(G\) the gravitational constant) indicates that we are sufficiently far from the gravitational source for the effects of general relativity to be negligible. Indeed, at great distances, the length \( \frac{2GM}{c^2} \) is completely negligible.}.\\

Consequently, it is possible to build the Schwarzschild interior metric associated with the population of negative masses by applying the same calculation scheme as for the population of positive masses, thus constituting the solution to the second field equation \ref{eq_positivemass2} as follows:
\begin{equation}
{\bar{\mathrm{d}s}}^{2} = \left[ \frac{3}{2} \sqrt{\left( 1 + \frac{{r_n}^{2}}{{\hat{r}}^{2}} \right)} - \frac{1}{2} \sqrt{\left( 1 + \frac{r^{2}}{{\hat{r}}^{2}} \right)} \right]^{2} \mathrm{d}x^{0^{2}} - \frac{\mathrm{d}r^{2}}{1 + \frac{r^{2}}{{\hat{r}}^{2}}} - r^{2} \left( \mathrm{d}\theta^{2} + \sin^{2}\theta \mathrm{d}\phi^{2} \right).
\label{chapter3_eq_tov7_2}
\end{equation}
This metric must join the Schwarzschild exterior metric:
		\begin{equation}
		{\bar{\mathrm{d}s}}^{2} = \left(1 + \frac{2 G M}{c^2 r}\right) c^2 \mathrm{d}x^{0^{2}}
		- \frac{\mathrm{d}r^2}{1 + \frac{2 G M}{c^2 r}} 
		- r^2 \left( \mathrm{d}\theta^2 + \sin^2\theta \mathrm{d}\phi^2 \right).
		\label{chapter3_eq_tov7_2_ext}
		\end{equation}
We can deduce that a particle with negative mass will undergo a repulsive gravitational field due to the effect of a distribution of positive masses.\\

The Janus model presents a new paradigm, extending general relativity by describing the universe as a four-dimensional manifold \( M_4 \), endowed with two distinct metrics. These metrics are solutions to the system of coupled field equations \eqref{chapter3_eq34} and \eqref{chapter3_eq35}.\\

Let's now consider the case, still in the Newtonian approximation, where the geometry is determined by the presence of negative mass, corresponding to regions of space dominated by negative masses, such as near the dipole repeller (\cite{Hoffman2017}).

\subsection{\texorpdfstring{Newtonian approximation of the field generated by a negative mass $\Mbar$}{Newtonian approximation of the field generated by a negative mass Mbar}}

In regions where negative masses dominate, such as near the dipole repeller (\cite{Hoffman2017}), the system becomes in mixed-mode form:
\begin{align}
\Rmixmunu - \frac{1}{2} \gmixmunu R &= \chi \sqrt{\frac{|\gbar|}{|g|}} \tmixmunu , \label{eq_negativemass1} \\
\Rbarmixmunu - \frac{1}{2} \gbarmixmunu \Rbar &= -\bar{\chi} \Tbarmixmunu . \label{eq_negativemass2}
\end{align}
If we consider the impact of the presence of negative masses on the geometry of spacetime structured by the metric tensor of the first field equation \ref{eq_negativemass1} associated with the population of positive masses, we can define the corresponding interaction tensor \ref{tenseur_dipole} as follows: 
\begin{equation}
{\mathcal{T}^{\nu}_{\mu}} = 
\begin{pmatrix}
\bar{\rho}\bar{c}^2 & 0 & 0 & 0 \\
0 & -\bar{p} & 0 & 0 \\
0 & 0 & -\bar{p} & 0 \\
0 & 0 & 0 & -\bar{p}
\end{pmatrix}.
\label{tenseur_dipole}
\end{equation}
Thus, the impact of the pressure gradient of negative masses on the geodesics followed by ordinary matter and positive-energy photons according to the field equation \ref{eq_negativemass1} translates into the following Tolman--Oppenheimer--Volkoff equation:
\begin{equation}
\frac{{{\bar{p}'}}}{{\bar{c}}^2} = -\frac{{m} - \frac{4\pi G {\bar{p}} r^3}{{\bar{c}}^4}}{r (r + 2{m})} \left( {\bar{\rho}} - \frac{{\bar{p}}}{{\bar{c}}^2} \right).
\label{chapter3_eq_tov6_d}
\end{equation}

Therefore, it is possible to build the Schwarzschild interior metric solution in this manner:
\begin{equation}
{\mathrm{d}s}^{2} = \left[ \frac{3}{2} \sqrt{\left( 1 + \frac{{r_n}^{2}}{{\hat{r}}^{2}} \right)} - \frac{1}{2} \sqrt{\left( 1 + \frac{r^{2}}{{\hat{r}}^{2}} \right)} \right]^{2} \mathrm{d}x^{0^{2}} - \frac{\mathrm{d}r^{2}}{1 + \frac{r^{2}}{{\hat{r}}^{2}}} - r^{2} \left( \mathrm{d}\theta^{2} + \sin^{2}\theta \mathrm{d}\phi^{2} \right).
\label{chapter3_eq_tov7_2_d}
\end{equation}
This metric can be connected to the Schwarzschild exterior metric:
		\begin{equation}
		{\mathrm{d}s}^{2} = \left(1 + \frac{2 G M}{c^2 r}\right) c^2 \mathrm{d}x^{0^{2}}
		- \frac{\mathrm{d}r^2}{1 + \frac{2 G M}{c^2 r}} 
		- r^2 \left( \mathrm{d}\theta^2 + \sin^2\theta \mathrm{d}\phi^2 \right).
		\label{chapter3_eq_tov7_2_ext_d}
		\end{equation}
We can deduce that a particle of ordinary matter will undergo a repulsive gravitational field due to the effect of a distribution of negative masses.\\

Then, when the source of the gravitational field of the second field equation (\ref{eq_negativemass2}) is created by a negative mass, we can freely define the following energy-momentum tensor as follows:
\begin{equation}
\Tbarmixmunu = 
\begin{pmatrix}
\bar{\rho}\bar{c}^2 & 0 & 0 & 0 \\
0 & \bar{p} & 0 & 0 \\
0 & 0 & \bar{p} & 0 \\
0 & 0 & 0 & \bar{p}
\end{pmatrix}.
\label{tenseur_energie_impulsion_hh}
\end{equation}
We can therefore deduce the following Tolman--Oppenheimer--Volkoff equation:
\begin{equation}
\frac{{{\bar{p}'}}}{\bar{c}^2} = -\frac{\overline{m} + \frac{4\pi G \bar{p} r^3}{{\bar{c}}^4}}{r (r - 2\overline{m})} \left( {\bar{\rho}} + \frac{\bar{p}}{{\bar{c}}^2} \right).
\label{chapter3_eq_tov6_d2}
\end{equation}
Hence, the interior Schwarzschild metric can be constructed as follows:
\begin{equation}
\bar{{\mathrm{d}s}}^{2} = \left[ \frac{3}{2} \sqrt{\left( 1 - \frac{\bar{{r_n}}^{2}}{{\hat{r}}^{2}} \right)} - \frac{1}{2} \sqrt{\left( 1 - \frac{r^{2}}{{\hat{r}}^{2}} \right)} \right]^{2} \mathrm{d}x^{0^{2}} - \frac{\mathrm{d}r^{2}}{1 - \frac{r^{2}}{{\hat{r}}^{2}}} - r^{2} \left( \mathrm{d}\theta^{2} + \sin^{2}\theta \mathrm{d}\phi^{2} \right).
\label{chapter3_eq_tov7_1_d2}
\end{equation}
This metric matches the exterior Schwarzschild metric:
		\begin{equation}
		\bar{{\mathrm{d}s}}^{2} = \left(1 - \frac{2 G \Mbar}{\bar{c}^2 r}\right) \bar{c}^2 \mathrm{d}x^{0^{2}}
		- \frac{\mathrm{d}r^2}{1 - \frac{2 G \Mbar}{\bar{c}^2 r}} 
		- r^2 \left( \mathrm{d}\theta^2 + \sin^2\theta \mathrm{d}\phi^2 \right).
		\label{chapter3_eq_tov7_1_ext_d2}		
		\end{equation}
We can deduce that a particle of negative mass will undergo an attractive gravitational field due to the effect of a distribution of negative masses.\\

Both solutions (\ref{chapter3_eq_tov6_d}) and (\ref{chapter3_eq_tov6_d2}) reduces to the Euler equation approximately equal to \( -\frac{G\Mbar(r)\bar{\rho}(r)}{r^2} \) in the Newtonian limit, reflecting hydrostatic equilibrium\footnote{Where the pressure at the center of this negative mass spheroid is balanced by the negative gravitational force depending on density and mass.}.\\

The form of these two source tensors satisfies the Bianchi identities. This would obviously not be the case if the negative mass were to fall outside of this framework. For that, there would need to exist neutron stars of negative mass. However, the characteristic time of evolution of conglomerates of negative mass, their \textit{"cooling time"}, exceeds the age of the universe. These spheroidal conglomerates cannot evolve, so the content of this negative spacetime will be limited to a mixture of negative mass anti-hydrogen and anti-helium. Since nucleosynthesis cannot occur, there can be no anti-galaxies or anti-stars, regardless of their mass (\cite{Petit2023}). Consequently, there cannot exist anti-neutron stars. \\

Moreover, in the case where this negative spacetime would generate hyperdense stars through an as-yet-unknown mechanism, it would then be necessary to reconsider the form of these tensors. However, the current configuration satisfies all currently available and potentially available observational data.\\

%Photons of positive energy emitted by sources located behind the Dipole Repeller will experience a significant decrease in their magnitude due to the negative gravitational lensing effect. These photons then freely traverse this vast void. The effect will be maximal when the photons brush past this spheroidal conglomerate, where the entirety of the mass must be taken into account. However, it will be negligible when these photons pass through the central neighborhood (see \vref{photon-negative-mass}).\\

%Thus, we predict that when a map is established by the JWST telescope, the invisible mass will manifest its presence by a brightness attenuation, not over the entire disk, but in a ring (see \vref{attenuation}).

After verifying the mathematical and physical consistency of the Janus model, we now turn to its predictive capabilities. One of the most striking predictions concerns the existence of large voids and structures such as the dipole repeller. The Janus model not only accounts for these features but also offers novel predictions regarding the effect of negative gravitational lensing on the magnitudes of background sources. In the following short section, we explore the observational signatures of this phenomenon, with a particular focus on the implications for the dipole repeller. 

\section{Dipole repeller prediction}

\begin{figure}
	\centering
	\includegraphics[width=9cm]{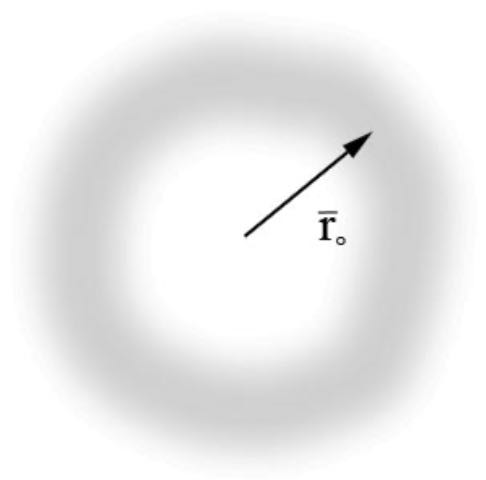}
	\caption{Attenuation of the magnitude of objects in the background of the dipole repeller.}
	\label{attenuation}
\end{figure}

The Janus model is essentially falsifiable in Popper's sense. It predicted a large-scale twin structure with large voids. This has been confirmed (\cite{Hoffman2017}). It predicted a very early birth of first-generation stars and galaxies. A new prediction this time concerns the magnitude of sources located in the background of the large void. According to the model, the magnitude of the light emitted by these distant sources will be attenuated by the negative gravitational lensing effect. This is a novel aspect, since it has been assumed that the two entities, positive and negative, interact only through antigravitation. Photons from these distant sources can then freely pass through the negative-mass conglomerates. This means that both external and internal geodesics must be used. The deflection effect of light rays will be greatest when they graze the surface of the object, with radius. This effect weakens as you move deeper into the object, becoming zero when the photons pass through its center (see \ref{photon-negative-mass}). Eventually, we will be able to map the magnitudes of objects in the background of the dipole repeller. Schematically, their luminosity will be attenuated in a ring-shaped pattern (see \vref{attenuation}).\\

This measurement will immediately give us the value of the radius $\rzbar$ of this formation.\\

After exploring the implications of the Janus model in the Newtonian approximation and its predictions for large-scale structures, such as the dipole repeller, we now move beyond these limitations. In a universe dominated by positive masses, certain astrophysical objects, such as neutron stars and supermassive black holes, exhibit strong gravitational effects that require a relativistic treatment. The following section addresses the challenges of extending the model to these extreme cases.

\section{Beyond the Newtonian approximation}

These objects are absent in the universe fold associated with negative masses. In our universe fold of ordinary matter, objects that deviate from the Newtonian approximation are neutron stars and hypermassive objects located at the center of galaxies, which early images show to be the seat of a strong gravitational redshift effect, darkening their central part. These objects are a priori manageable using the classic pair of outer and inner metrics, taking rotation into account. It should be remembered that we are under no obligation to provide the form of the source tensor of the other sector, in this case an interaction tensor, whose form would be precisely imposed by the Bianchi identities. It's conceivable that one day someone will provide the exact form of this tensor.\\

But even in the absence of such an object, there is no a priori inconsistency.

\section{Conclusion}

The genesis of the Janus model spanned several decades. The starting point, in 1967, was Andreï Sakharov's attempt to provide an initial explanation for the absence of observations of primordial antimatter, which remains a significant flaw in the Standard Model $\Lambda$CDM. This model offers no explanation for the loss of half of the universe's content. Sakharov therefore proposed a universe structure with two sectors, the second being T-symmetrical to our own. A few years later, in 1970, through the application of symplectic geometry, mathematician Jean-Marie Souriau demonstrated that this inversion of the time coordinate, i.e., T-symmetry, is synonymous with the inversion of energy and mass. Pushing this idea of global symmetry further, Sakharov envisioned a twin universe that is CPT-symmetrical to ours. In this scenario, the invisible components of the universe reduce to negative-mass antimatter.\\

In 1994, we proposed that this universe structure corresponds to a two-fold cover of a projective $\mathbb{P}^4$, by a compact universe with the topology of a $\mathbb{S}^4$ sphere. The two singularities of this spherical universe, the Big Bang and the Big Crunch, then coincide. By introducing a tubular structure, these singularities disappear. This configuration consists of two PT-symmetrical folds. These adjacent sectors are assumed to interact solely through gravity. Therefore, the interaction between positive masses in one sector and negative masses in the other sector must be taken into account. \\

However, the introduction of negative masses is not feasible within the framework of general relativity, as it would result in interaction laws that are incompatible with known physical principles. Thus, a bimetric model is proposed. A system of coupled field equations is then constructed from an action, whose form eliminates the problematic runaway effect. The interaction laws in the model dictate that masses of the same sign attract each other according to Newton’s law, while masses of opposite signs repel each other following an anti-Newtonian law. Since these masses are mutually exclusive, the negative mass can be neglected in the vicinity of the Sun, and the first field equation then aligns with Einstein's equation. \\

In this way, the model remains consistent with local relativistic observations, such as the advance of Mercury's perihelion and the deflection of light by the Sun. Therefore, the Janus model can be considered an extension of general relativity. An exact, time-dependent solution is constructed, revealing a generalized energy conservation law that applies to both sectors. When adapting the model to observations, it becomes evident that an accelerating expansion is present, imposing a fundamental dissymmetry between the two entities involved. \\

In this framework, the vast majority of negative mass replaces the hypothetical components of dark matter and dark energy. As a result, the matter distribution is approximately 5\% visible matter and 95\% negative mass, which is invisible because it emits photons of negative energy that elude detection by our observation instruments. This dissymmetry implies that, following decoupling, the negative masses form a regular network of spheroidal conglomerates, while the positive mass, confined to the remaining space, adopts a patchy distribution. \\

The model also accounts for the existence of large voids, with the dipole repeller being the first identified among them. At the centers of these large voids are invisible spheroidal conglomerates that behave like giant protostars, with cooling times exceeding the age of the universe. These objects, which emit negative-energy photons corresponding to light in the red and infrared regions, do not evolve and do not give rise to stars, galaxies, or atoms heavier than helium. Life, therefore, is absent from this negative sector, which consists of a mixture of negative-mass antihydrogen and antihelium.\\

Furthermore, the model explains the very early formation of first-generation stars and galaxies, as recently demonstrated by the James Webb Space Telescope. We then examine the issue of the model's mathematical consistency, specifically whether the Bianchi identities are satisfied. We show that they can be asymptotically satisfied under conditions corresponding to the Newtonian approximation. \\

Lastly, we address the question of objects that do not fit within this approximation, primarily located on the positive-mass side. We assert that we are not required to provide the exact form of the interaction tensor in such cases, as it is determined by the zero-divergence condition. The lack of definition of this tensor does not invalidate the consistency of a non-linear solution.

\section{Author contributions}
Jean-Pierre Petit designed the model, developped the theoretical formalism, drafted the manuscript and prepared the figures. Florent Margnat managed the submission and peer-review process. Hicham Zejli corrected many typos from the draft and improved the layout of the manuscript.

\newpage
\printbibliography[heading=bibintoc]

\end{document}